\documentclass[aoas,preprint]{imsart}
\RequirePackage[OT1]{fontenc}
\RequirePackage{amsthm,amsmath}
\RequirePackage{natbib}
\RequirePackage[colorlinks,citecolor=blue,urlcolor=blue]{hyperref}
\usepackage{xr}
\usepackage[figuresright]{rotating}
\usepackage{amssymb,amsmath,booktabs,multirow,array}
\usepackage{mathrsfs, bm}
\usepackage{algorithm}
\usepackage{algorithmic}
\usepackage{float}

\def\iid{\stackrel{\mathrm{iid}}{\sim}}

\def\bm{\boldsymbol}
\def\bz{\bm z}
\def\by{\bm y}
\def\bx{\bm x}
\def\bs{\bm s}
\def\bA{\bm A}
\def\bSigma{\bm \Sigma}

\def\hTh{\hat{\Theta}}
\def\bbeta{\bm \beta}
\def\be{\bm e}

\def\bby{ \mathbf{ y}}
\def\bbs{\textsf{\textbf{s}}}

\def\bU{\bm U}
\def\bR{\bm R}
\def\vect{\mbox{vec}}

\DeclareMathOperator*{\argmax}{\arg\!\max}
\theoremstyle{plain}
\newtheorem{theorem}{Theorem}
\newtheorem{lemma}{Lemma}
\arxiv{arXiv:1402.4239}

\begin{document}

\begin{frontmatter}
\title{Modeling Covariate Effects in Group Independent Component Analysis With Applications to Functional Magnetic Resonance Imaging}
\runtitle{Modeling covariate effects in group ICA}

\begin{aug}
\author{\fnms{} \snm{Ran Shi}\thanksref{}\ead[label=e1]{rshi3@emory.edu}}
\and
\author{\fnms{} \snm{Ying Guo}\thanksref{t1}\ead[label=e2]{yguo2@emory.edu}}

\thankstext{t1}{Supported by NIH grants 1R01MH105561-01 and 5R01MH079448-05, and a URC/ACTSI grant funded by Emory University Research Committee.}

\runauthor{R. SHI AND Y. GUO}

\affiliation{Emory University}

\address{Department of Biostatistics and Bioinformatics\\
Rollins School of Public Health\\
Emory University\\
1518 Clifton Rd.\\
Atlanta, Georgia 30322 USA\\
\printead{e1}\\
\phantom{E-mail:\ }\printead*{e2}}

\end{aug}

\begin{abstract}
	Human brains perform tasks via complex functional networks consisting of separated brain regions. A popular approach to characterize brain functional networks in fMRI studies is independent component analysis (ICA), which is a powerful method to reconstruct latent source signals from their linear mixtures. An important goal in many fMRI studies is to investigate how clinical and demographic variables affect brain functional networks. Existing ICA methods, however, cannot directly incorporate these covariate effects in ICA decomposition. Hence, researchers can only address this need via heuristic post-ICA analyses which may be inaccurate and inefficient. In this paper, we propose a hierarchical covariate ICA (hc-ICA) model that provides a formal statistical framework for estimating and testing covariate effects in ICA. To obtain the maximum likelihood estimates of hc-ICA, we first present an exact EM algorithm with analytically tractable E-step and M-step. We then develop a subspace-based approximate EM that can significantly reduce computation time while retaining high estimation accuracy. To test covariate effects on functional networks, we introduce a voxel-wise approximate inference procedure which eliminates the need of computationally expensive covariance estimation and inversion. We demonstrate the advantages of our methods over the existing method via simulation studies. The proposed hc-ICA is applied to an fMRI study to examine the effects of Zen meditation practice on brain functional networks. The results show that meditators demonstrate better synergy or functional connectivity in several relevant brain functional networks as compared with the control. These findings were not revealed in previous analyses of this data using existing group ICA methods.
\end{abstract}

\begin{keyword}
\kwd{Blind source separation}
\kwd{Brain functional networks}
\kwd{Mixture of Gaussian distributions}
\kwd{EM algorithm}
\kwd{Subspace concentration}
\end{keyword}

\end{frontmatter}

\section[]{Introduction}
Functional magnetic resonance imaging (fMRI) is one of the most commonly used imaging technologies to investigate neural activities in the brain. In fMRI studies, the observed data represent combined signals generated from various brain functional networks (BFNs). Each of these networks consists of a set of spatially disjoint brain regions that demonstrate similar temporal patterns in blood oxygenation level dependent (BOLD) fMRI signals. One major goal in fMRI analysis is to identify these underlying functional networks and characterize their spatial distributions as well as temporal dynamics. Independent component analysis (ICA) has become one of the most commonly used tools to achieve this goal in neuroimaging studies. As a special case of blind source separation, ICA aims to separate observed data into a linear combination of latent components that are statistically independent. ICA a fully data-driven approach that does not require any a priori information about the underlying source signals. ICA was initially applied to analyze single-subject fMRI data \citep{mckeown1997analysis, biswal1999blind, calhoun2001method, beckmann2005tensorial, lee2011independent}. Denote by $\bm Y$ the $T\times V$ fMRI data matrix for one subject, where $T$ is the number of fMRI scans and $V$ is the number of voxels in the 3D brain image acquired during each scan. That is, each row of $\bm Y$ represents a concatenated 3D image. Classical noise-free ICA model can be applied to decompose the observed fMRI data as $\label{eq:ICA}\bm Y_{T\times V} = \bm A_{T\times q}\bm S_{q\times V}$, where $q$ is the total number of source signals. Each row of $\bm S$ represents a concatenated 3D map of a spatial source signal. $\bm A$ is the temporal mixing matrix which mixes the $q$ spatial sources to generate the observed time series of fMRI images. The $q$ source signals are assumed to be statistically independent and hence are called independent components (ICs). For fMRI data, each row of $\bm S$ and the corresponding column of $\bm A$ represent the spatial distribution and temporal dynamics for a BFN. And statistical independence is usually assumed in the spatial domain for fMRI, i.e. the rows in $\bm S$ are independent.

To decompose multi-subject fMRI data, ICA has been extended for group analysis, which is referred to as group ICA (Calhoun et al., 2001). One commonly used group ICA framework in fMRI analysis is the temporal concatenation group ICA (TC-GICA). In TC-GICA, the $T\times V$ fMRI data matrices from $N$ subjects are stacked on the temporal domain to form a tall $TN\times V$ group data matrix. The concatenated group data is then decomposed into the product of a $TN\times q$ group mixing matrix and a $q\times V$ spatial source matrix with independent rows. Many of the existing group ICA methods \citep{calhoun2001method, beckmann2005tensorial, guo2008unified, guo2011general} were developed under the TC-GICA framework. A notable restriction of the TC-GICA framework is the assumption on homogeneous spatial source signals across subjects. To relax this restriction, \citet{guo2013hierarchical} proposed a hierarchical group ICA (H-GICA) model to accommodate between-subject variability in spatial source signals by incorporating subject-specific random effects in ICA.

In recent years, the field of neuroimaging has been moving toward a more network-oriented view of brain function. Existing neuroscience literature has provided evidence that BFNs can vary considerably with subjects' clinical, biological and demographic characteristics. For example, neuroimaging studies have shown that neural activity and connectivity in specific functional networks are significantly associated with pathophysiology of mental disorders and their responses to treatment \citep{anand2005activity,greicius2007resting,chen2007brain,sheline2009default}. Other studies have found activities patterns in major functional networks vary with subject demographic factors including age and gender \citep{quiton2007sex,cole2010age}. Given these early findings, there is a strong need to formally and systematically quantify the effects of subjects' characteristics on the distributed patterns of BFNs and to accurately evaluate the differences in BFNs between subject groups (e.g. diseased v.s. normal) while accounting for potential confounding factors. Our method development in this paper aims to address this need by providing a formal statistical framework to model covariate effects on BFNs via group ICA.

One limitation of the existing group ICA methods is that they do not incorporate subjects' covariate information in ICA decomposition. Currently, the covariate effects in ICA are assessed with two kinds of heuristic approaches. The first approach is through conducting single-subject ICA separately on each subject's data, selecting matched ICs from each subject and then performing group analysis on the selected subject-level IC maps \citep{greicius2007resting}. A major problem with this approach is that it is often challenging to identify matching ICs across subjects since ICA results are only identifiable up to a permutation of the source signals. Furthermore, since most ICA algorithms are stochastic, ICs extracted from separate ICA runs for different subjects are often not comparable to each other. A more advanced approach for covariates effects is two-stage analysis based on TC-GICA. The back-construction in \citet{calhoun2001method} and the dual regression in \citet{beckmann2009group} are two representative methods in this category. To be specific, these methods first perform TC-GICA to extract common IC maps at the group level and then reconstruct subject-specific IC maps by performing post-ICA analysis. Then the covariate effects are evaluated via secondary tests or regressions on the subject-specific maps. These methods do not take into account the random variabilities introduced in reconstructing subject-specific IC maps. Estimating and testing the covariate effects on IC maps can lead to loss of accuracy and efficiency.

In this paper, we propose a new \emph{h}ierarchical \emph{c}ovariate \emph{ICA} model (hc-ICA) that directly incorporates covariate effects in group ICA decomposition. The hc-ICA model first decomposes each subject's fMRI data into linear mixtures of subject-specific spatial source signals (ICs). The subject-specific ICs are then modeled in terms of population-level source signals, covariate effects and between-subject random variabilities. To the best of our knowledge, hc-ICA is the first model-based group ICA method for modeling covariate effects on brain functional networks. By formally accounting for covariates in ICA decomposition, hc-ICA overcomes the aforementioned issues in the existing heuristic approaches and provides a more accurate and efficient method to estimate and test covariate effects on the ICs. Furthermore, hc-ICA can provide model-based prediction of distributed patterns of brain functional networks for various clinical or demographic subgroups.

Our hc-ICA model is developed under the hierarchical probabilistic ICA modeling framework that is first proposed in \citet{guo2013hierarchical}. The current work provides several important contributions that can lead to major advancements in methods of hierarchical ICA. First, the proposed hc-ICA allows us to examine how subjects' covariates help explain the between-subject variability in ICs. Secondly, we propose a novel subspace-based EM algorithm for obtaining maximum likelihood estimates of parameters in the hierarchical ICA model. The proposed new EM exploits the biological characteristics of fMRI source signals to achieve a significant reduction in computational time while retain high accuracy in model estimation. By focusing on a subspace of the latent states of source signals, the new EM algorithm achieves a computation load that scales linearly with the number of ICs, which is significantly more efficient as compared with the exponential growth of computing time in the original EM algorithm proposed for hierarchical ICA \citet{guo2013hierarchical}. We provide theoretical justification for the subspace-based EM in imaging analysis and also present empirical evidence through simulations that demonstrate the subspace-based EM can provide highly accurate approximation. The proposed approximation method can potentially be generalized to develop algorithms with polynomial complexity for other big imaging data with sparse signals. Thirdly, we propose a voxel-wise approximate inference procedure to test covariate effects in hc-ICA, which eliminates the need for computationally expensive estimation of the huge covariance matrix. Results from simulation studies show that our proposed hc-ICA method has better performance than the existing TC-GICA two-stage method in terms of both estimation accuracy and statistical power.

Our motivating data example is an fMRI experiment to compare the spatio-temporal differences in BFNs between Zen meditators and the control. Previously, \citet{guo2008unified} and \citet{guo2011general} analyzed this data example under the TC-GICA framework and focused on examining the temporal differences in the BFNs while assuming the BFNs have homogeneous spatial distributions across subjects. Recently,  we applied the random effects ICA model in \citet{guo2013hierarchical} to this data to accommodate between-subject variabilities in the spatial domain. In this paper, the proposed hc-ICA method can, for the first time, conduct formal statistical estimation and inference for between-group differences in the spatial distributed patterns of BFNs and also provide model-based BFN maps for each group. We illustrate the results from hc-ICA for two relevant BFNs: the task-related network and the default mode network. Findings from hc-ICA revealed statistically significant differences in the spatial distributions of these two BFNs between the two groups. Specifically, our results show that meditators demonstrate better synergy or connectivity in both networks, indicating the meditators have more regularized neural activity in BFNs. We also analyzed the Zen meditation data using the existing TC-GICA two-stage method, which failed to detect some important between-group differences in the networks.

The rest of this paper is organized as follows. Section \ref{sec2} introduces the hc-ICA framework including data preprocessing, model building, estimation and inference. Section \ref{sec3} reports simulation results for comparing hc-ICA to the existing heuristic method, comparing the subspace-based EM to the exact EM algorithms and comparing the proposed inference method to the two-stage approach for testing covariate effects. Section \ref{sec4} focuses on detailed analysis of the Zen meditation data. Conclusions and discussions are presented in Section \ref{sec5}. The web supplementary materials documented the derivations and proofs for the algorithms, as well as more details and findings in real data analysis.

\section[]{Methods}\label{sec2}
This section introduces the hc-ICA framework, which includes the preprocessing step, the hc-ICA model, estimation algorithms and the inference procedure.
\subsection{Preprocessing prior to ICA}
Prior to an ICA algorithm, some preprocessing steps such as centering, dimension
reduction and whitening of the observed data are usually
performed to facilitate the subsequent ICA decomposition \citep{hyvarinen2001independent}. Here, we present the preprocessing procedure prior to the hc-ICA. Suppose that the fMRI study consists of $N$ subjects. For each subject, the fMRI signal are acquired at $T$ time points across $V$ voxels. Let $\widetilde{\by_i}(v) \in \mathbb{R}^T$ be the centered time series recorded for subject $i$ at voxel $v$. Then $\widetilde{\bm  Y}_i = [\widetilde{\by_i}(1),...,\widetilde{\by_i}(V)]$ is the $T\times V$ fMRI data matrix for subject $i$.

Under the paradigm of group ICA, we perform the following dimension reduction and whitening procedure on the original fMRI data: for $i=1,...,N$,
\begin{equation}\label{eq:preproc} \bm Y_i= (\bm\Lambda_{i, q}-\tilde{\sigma}^2_{i, q}\bm I_q)^{-\frac{1}{2}}\bm U'_{i, q}\widetilde{\bm Y}_i,
\end{equation}
where $\bm U_{i, q}$ and $\bm \Lambda_{i, q}$ contain the first q eigenvectors and eigenvalues
based on the singular value decomposition of $\bm Y_i$.  The residual variance, $\tilde{\sigma}^2_{i, q}$, is the average of the smallest $T-q$ eigenvalues that are not included in $\bm \Lambda_{i, q}$ representing the variability in $\bm Y_i$ that is not accounted by the first $q$ components. The parameter $q$, which is the number of ICs, can be determined using the Laplace approximation method \citep{minka2000automatic}. Throughout the rest of our paper, we will present the model and methodologies based on the preprocessed data $\bm Y_i = [\by_i(1),...,\by_i(V)]$  ($i=1,...,N$), which are $q\times V$ matrices.
\subsection{A hierarchical covariate ICA model (hc-ICA)}
In this section, we present a hierarchical covariate ICA (hc-ICA) model for evaluating covariate effects on brain functional networks using multi-subject fMRI data. The first-level model of hc-ICA decomposes a subject's observed fMRI signals into a product of subject-specific spatial source signals and a temporal mixing matrix to capture between-subject variabilities in the spatio-temporal processes in the functional networks. We include a noise term in this ICA model to account for residual variabilities in the fMRI data that are not explained by the extracted ICs, which is known as probabilistic ICA \citep{beckmann2004probabilistic}. To be specific, the first-level of hc-ICA is defined as,
\begin{equation}
\label{eq:level1}
\bm y_{i}(v) = \bA_{i}\bm s_{i}(v) +\bm e_i(v),
\end{equation}
where $\bs_i(v)=[s_{i1}(v),...,s_{iq}(v)]'$ is a $q\times 1$ vector with $s_{i\ell}(v)$ representing the spatial source signal of the $l$th IC (i.e., functional network or source signal) at voxel $v$. The $q$ elements of $\bs_i(v)$ are assumed to be independent and non-Gaussian. $\bA_{i}$ is the $q\times q$ mixing matrix for subject $i$ which mixes$\bs_i(v)$ to generate the observed fMRI data. Since $\bm Y_i$ is whitened data, it can be shown that the mixing matrix, $\bA_i$, is orthogonal \citep{hyvarinen2000independent}. $\bm e_i(v)$ is a $q\times 1$ vector that represents the noise in the subject's data and $\bm e_i(v)\sim  \mbox{N}(\bm 0, \bm E_v)$ for $v=1,...,V$. Since the spatial variabilities and correlations among $\bm y_{i}(v)$ across voxels are modeled by the spatial source signals, we assume $\bm e_i(v)$ are independent across voxels with spatial stationarity in their variance, i.e., $\bm E_v=\bm E$ for $v=1,...,V$. Prior to ICA, preliminary analysis such as pre-whitening \citep{bullmore1996statistical} can be performed to remove temporal correlations in the noise term and to standardize the variability across voxels. Therefore, we follow previous work \citep{beckmann2004probabilistic, beckmann2005tensorial, guo2008unified, guo2011general} and further assume the covariance for the noise term is isotropic, i.e. $\bm E=\nu^2_0 \bm I_q$.

At the second-level of hc-ICA, we further model subject-specific spatial source signals $\bs_i(v)$ as a combination of the population-level source signals, the covariate effects and additional between-subject random variabilities:
\begin{equation}\label{eq:level2}
\bs_i(v) = \bs_0(v) + \bbeta(v)^{\prime}\bm x_{i} + \bm \gamma_{i}(v),
\end{equation}
where $\bs_0(v)=[s_{01}(v),...,s_{0q}(v)]'$ is the population-level spatial source signals of the $q$ statistically independent and non-Gaussian ICs; $\bm x_i =[x_{i1},...,x_{ip}]'$ is the $p\times 1$ covariate vector containing subject-specific characteristics such as the treatment or disease group, demographic variables and biological traits; $\bbeta(v)$ is a $p\times q$ matrix where the element $\beta_{k\ell}(v)$ $(k=1,...p, \ell=1,...,q)$ in $\bbeta(v)$ captures the effect of the $k$th covariate on $\ell$ th IC at voxel $v$; $\bm \gamma_{i}(v)$ is a $q\times 1$ vector reflecting the random variabilities among subjects after adjusting the covariate effects. We assume $\bm \gamma_{i}(v)\iid\mbox{N}(\bm 0, \bm D)$ where $\bm D= \mbox{diag}(\nu_1^2,...,\nu_q^2)$. IC-specific variances specified in $\bm D$ allow us to accommodate different levels of between-subject variability across ICs. By formally modeling covariate effects in ICA, the proposed hc-ICA model can provide model-based estimates of spatial distributions of BFNs for subgroups defined by covariates. It can also allow us to assess the adjusted effects of primary covariates, such as disease or treatment group, on functional networks while controlling for potential confounding factors. This provides important benefits for understanding of neural basis of diseases such as mental disorders which are known to be affected by many other demographic and clinical factors \citep{quiton2007sex, greicius2007resting, cullen2009preliminary, cole2010age}.

\subsection{Source signal distribution assumptions}
Following \citet{guo2011general, guo2013hierarchical}, we choose a mixture of Gaussian distributions (MoG) as our source distribution model for the population-level spatial source signals, $\bs_0(v)$, in \eqref{eq:level2}. MoG has several desirable properties for modeling fMRI signals. First, within each BFN, only a small percentage of locations in the brain are activated whereas most brain areas exhibit background fluctuations \citet{biswal1999blind}. MoG is well suited to model such mixed patterns. In addition, MoG can captures various types of non-Gaussian signals \citep{xu1997maximum, kostantinos2000gaussian} and also offer tractable likelihood-based estimations \citep{mclachlan2004finite}.

Specifically, for $\ell=1,\ldots,q$ we assume that
\begin{equation}\label{eq:level31} s_{0\ell}(v) \sim \mbox{MoG}(\bm \pi_{\ell}, \bm \mu_{\ell}, \bm \sigma^2_{\ell}), \qquad v=1,...,V,\end{equation}
where $\bm \pi_{\ell}=[\pi_{\ell,1},...,\pi_{\ell,m}]'$ with $\sum_{j=1}^m \pi_{\ell, j}=1$, $\bm \mu_{\ell}=[\mu_{\ell,1},...,\mu_{\ell,m}]'$ and $ \bm \sigma^2_{\ell} = [\sigma^2_{\ell,1},...,\sigma^2_{\ell,m}]'$; $m$ is the number of Gaussian components in MoG. The probability density of $\mbox{MoG}(\bm \pi_{\ell}, \bm \mu_{\ell}, \bm \sigma^2_{\ell})$ is $\sum_{j=1}^m \pi_{\ell, j} g(s_{0\ell}(v); \mu_{\ell,j}, \sigma^2_{\ell, j})$ where $g(\cdot)$ is the pdf of the (multivariate) Gaussian distribution. In fMRI applications, mixtures of two to three Gaussian components are sufficient to capture the distribution of fMRI spatial signals, with the different Gaussian components corresponding to the background fluctuation and the negative or positive fMRI BOLD effects respectively \citep{beckmann2004probabilistic, guo2008unified, guo2011general}. In our model, we interpret $j=1$ as the background fluctuation state by default throughout the rest of the paper.

To facilitate derivations in models involving MoG, latent states variables are often used \citep{mclachlan2004finite}. Here we define latent states $\bz(v)=[z_{1}(v),...,z_q(v)]'$ at voxel $v$ as follows. For $\ell=1,...,q$,  $z_{\ell}(v)$ takes value in $\{1,\ldots,m\}$ with probability $p[z_{\ell}(v)=j]=\pi_{\ell,j}$ for $j=1,..,m$. Conditional on $\bz(v)$, we can rewrite our source distribution model as,
\begin{equation}\label{eq:level32}
\bm s_0(v) = \bm \mu_{ \bz(v)} + \bm \psi_{\bz(v)},
\end{equation}
where $\bm\mu_{ \bz(v)} = [\mu_{1,z_1(v)},...,\mu_{q,z_q(v)}]'$ and $ \bm \psi_{\bz(v)}=[\psi_{1, z_1(v)},...,\psi_{q, z_q(v)}]'$; $\bm \psi_{\bz(v)} \sim \mbox{N}(\bm 0,  \bSigma_{\bz(v)})$ with $\bSigma_{\bz(v)}$ $= \mbox{diag}(\sigma^2_{1,z_1(v)},...,\sigma^2_{q,z_q(v)})$.

\subsection{Maximum likelihood estimation}
We propose to estimate parameters in the hc-ICA model through a maximum likelihood (ML) approach using the EM algorithm. Based on models in \eqref{eq:level1}, \eqref{eq:level2} and \eqref{eq:level32}, the complete data log-likelihood for our model is
\begin{equation}l(\Theta ; \mathcal{Y},  \mathcal{X}, \mathcal{S}, \mathcal{Z})=\sum_{v=1}^V l_v(\Theta ; \mathcal{Y},  \mathcal{X}, \mathcal{S}, \mathcal{Z}),\end{equation}
where $\mathcal{Y}=\{ \bm y_{i}(v): i=1,...,N; v=1,\ldots,V\}$, $\mathcal{X}=\{\bm x_i: i=1,...,N\}$, $\mathcal{S}=\{\bm \bs_i(v): i=0,...,N, v=1,...,V\}$ and $\mathcal{Z}=\{\bm z(v): v=1,...,V\}$; the parameters are $\Theta =\{\{\bbeta(v)\}, \{\bm A_i\}, \bm E, \bm D, \{\bm \pi_{\ell}\}, \{\bm\mu_{\ell}\}, \{\bm \sigma^2_{\ell}\}: i=1,...,N, v=1,...,V, \ell=1,...,m\}$. The detailed expressions for the complete data log-likelihood function at each voxel $v$ is:
\begin{align}\label{likv}
\displaystyle l_v(\Theta ; \mathcal{Y},  \mathcal{X}, \mathcal{S}, \mathcal{Z})
&= \sum_{i=1}^N\bigg[\log g\left(\by_i(v); \bA_i\bs_i(v), \bm E\right) + \log g\left(\bs_i(v) ; \bs_0(v)+\bbeta(v)'\bm x_i, \bm D\right)  \bigg] \nonumber\\
&+ \log g\left(\bs_0(v); \bm \mu_{\bz(v)}, \bSigma_{\bz(v)}\right) + \sum_{\ell=1}^q \log \pi_{l, \bz_l(v)}.
\end{align}

\subsubsection{The exact EM algorithm}
 An exact EM which has an explicit E-step and M-step are introduced in this section to obtain ML estimates for the parameters in hc-ICA.

\textbf{E-step}: In the E-step, given the parameter estimates $\hTh^{(k)}$ from the last step, we derive the conditional expectation of the complete data log-likelihood  given the observed data as follows:
\begin{equation}\label{qfun}
Q(\Theta|\hTh^{(k)})=\sum_{v=1}^V E_{\bbs(v), \bz(v) \mid \mathbf{y}(v)}\left[l_v(\Theta ; \mathcal{Y},  \mathcal{X}, \mathcal{S}, \mathcal{Z})\right],
\end{equation}
where $\mathbf{y}(v) = [\by_1(v)',...,\by_N(v)']'$ represents the group data vector from the $N$ subjects at voxel $v$, $\bbs(v) =[\bm s_1(v)',..., \bm s_N(v)',\bm s_0(v)']'$ is the vector containing latent source signals on both the population and individual level. The detailed definition of $Q(\Theta|\hTh^{(k)})$ is available in the section 1 of web supplementary materials. The evaluation of $Q(\Theta|\hTh^{(k)})$ relies on obtaining $p\left[\bbs(v), \bz(v) |\mathbf{y}(v); \hTh^{(k)}\right]$ as well as its marginal distributions, which consists of the following three steps. First, we determine $p\left[\bbs(v) |\bz(v), \mathbf{y}(v); \hTh^{(k)}\right]$, which is a multivariate Gaussian distribution. Second, we evaluate the probability mass functions, $p\left[\bz(v) | \mathbf{y}(v); \hTh^{(k)}\right]$ through an application of the Bayes's Theorem. We finally obtain $p\left[\bbs(v)|\mathbf{y}(v); \hTh^{(k)}\right]$ by convolving the distributions derived in the previous two steps. More details can be found in section 2 of the supplementary material.

Given these probability distributions, we can derive the analytical forms for the conditional expectation in \eqref{qfun}. For illustration purpose, two main quantities of interest in \eqref{qfun} are given as follows:
\begin{eqnarray}
E[\bbs(v)\mid \bby(v);\Theta] &= &\sum_{\bz(v)\in \mathcal R}p[\bz(v) \mid \bby(v);\Theta] E[\bbs(v)\mid \bby(v), \bz(v); \Theta],\nonumber\\
E[\bbs(v)^{\otimes 2}\mid \bby(v);\Theta] &=& \sum_{\bz(v)\in \mathcal R}p[\bz(v) \mid \bby(v);\Theta] E[\bbs(v)\mid \bby(v), \bz(v); \Theta]^{\otimes 2}\nonumber\\
& &+\sum_{\bz(v)\in \mathcal R}p[\bz(v) \mid \bby(v);\Theta]\mbox{Var}[\bbs(v)\mid \bby(v), \bz(v); \Theta],\nonumber
\end{eqnarray}
where $\mathcal{R}$ represents the set of all possible values of $\bz(v) $, i.e. $\mathcal{R}=\{\bz^r\}_{r=1}^{m^q}$ where $\bz^r = [z^r_1,...,z^r_q]'$ and $z^r_\ell\in\{1,...,m\}$ for $\ell=1,...q$; the notation $\bm a^{\otimes 2}$ for a vector $\bm a$ stands for $\bm a\bm a'$.

Based on the results presented above, our E-step is fully tractable without the need for iterative numerical integrations.

\textbf{M-step}: In the M-step, we update the current parameters estimates $\hTh^{(k)}$ to
\begin{equation}\hTh^{(k+1)}=\argmax_{\Theta}Q(\Theta|\hTh^{(k)}).\end{equation}
 We have derived explicit formulas for all parameter updates. The updating rules are provided in section 3 of our supplementary material.

The estimation procedure for the exact EM algorithm is summarized in Algorithm 1. See section 1-3 of the supplementary material for details. After obtaining $\hTh$, we can estimate the population- and individual-level source signals and their variability based on the mean and variance of their conditional distributions. In fMRI analysis, researchers are often interested in thresholded IC maps to identify ``significantly activated" voxels in each BFN. Following previous work \citep{guo2011general, guo2013hierarchical}, we propose a thresholding method based on the mixture distributions for this purpose (section 6 of the supplementary material).
\begin{algorithm}
   \caption{The Exact EM Algorithm}
   \label{alg:example}
\begin{algorithmic}
   \STATE {\bfseries Initial values}:  Start with initial values $\hTh^{(0)}$ which can be obtained based on estimates from existing group ICA software.
   \REPEAT
     \STATE \textbf{E-step:}

        \STATE 1. Determine $p[\bbs(v) , \bz(v)\mid \bby(v);\hTh^{(k)}]$ and its marginals using the proposed three-step approach:
        \STATE \quad 1.a Evaluate the multivariate Gaussian $p[\bbs(v) \mid \bby(v), \bz(v); \hTh^{(k)}]$;
        \STATE \quad 1.b Evaluate $p[\bz(v) \mid \bby(v); \hTh^{(k)}]$;
        \STATE \quad 1.c $p[\bbs(v) , \bz(v)\mid \bby(v), \hTh^{(k)}] = p[\bbs(v) \mid \bby(v), \bz(v); \hTh^{(k)}]\times p[\bz(v) \mid \bby(v); \hTh^{(k)}]$;\\ $\quad\quad \ \ p[\bbs(v)\mid \bby(v), \hTh^{(k)}]=\sum_{\bz(v)\in \mathcal{R}}p[\bbs(v) , \bz(v)\mid \bby(v), \hTh^{(k)}] $;
     \STATE 2. Evaluate conditional expectations in $Q(\Theta|\hTh^{(k)}).$
     \STATE \textbf{M-step:}
     \STATE Update $\bbeta(v)$, $\bA_i$, $\pi_{\ell, j}$, $\mu_{\ell, j}$, $\sigma^2_{\ell, j}$;
     \STATE Update the variance parameters $\bm D, \bm E$.
\UNTIL{$\frac{\|\hTh^{(k+1)}-\hTh^{(k)}\|}{\|\hTh^{(k)}\|} <\epsilon$}
\end{algorithmic}
\end{algorithm}

\subsubsection{The approximate EM algorithm}
One major limitation of the exact EM algorithm is that its complexity increases exponentially with regard to the number of ICs. Specifically, $\mathcal{O}(m^q)$ operations are required for the exact EM algorithm to complete. The main reason is that, at each voxel, the exact EM evaluates and sums the conditional distributions across the whole sample space $\mathcal R$ of the latent state variables $\bz(v)$, which has a cardinality of $m^q$. A standard way to alleviate this issue is through mean field variational approximation. This method has been used by \citet{attias1999independent, attias2000variational} for single subject ICA and by \citet{guo2011general} for TC-GICA. However, the variational method cannot be easily generalized to other models such as hierarchical ICA because the derivation of the variational approximate distributions depends heavily on the model specifications. In most cases, the estimates for the variational parameters do not have analytically tractable expressions and require extra numerical iterations to obtain, which sometimes causes convergence problems.

In this section, we propose a new approximate EM algorithm for solving MoG-based ICA models in fMRI studies. Compared with the exact EM that needs $\mathcal{O}(m^q)$ operations, this new EM algorithm only requires $\mathcal{O}(mq)$ operations. The key idea behind the approximate algorithm is that instead of considering the whole sample space $\mathcal{R}$ of the latent state vector $\bz(v)$, we only focus on a small subspace of $\mathcal{R}$ in the algorithm. Theorem \ref{thm1} provides the definition for the subspace and shows that under certain conditions, the distribution of the latent state vectors is concentrated to the proposed subspace.

\begin{theorem}\label{thm1}
 Define $\mathcal{R}=\{\bz^r=[z^r_1,...,z^r_q]': z^r_\ell=j \ with\  j\in \{1,...,m\}, \ell=1,...,q \}$ for $r=1,...,m^q$, which is the domain of $\bz(v)$. For all $\bz(v)\in \mathcal R$, suppose that $p[z_\ell(v)=j]=\pi_{\ell,j}$ and that $p[\bz_\ell(v)=\bz^r_\ell]=\prod_{\ell=1}^q \pi_{\ell,z^r_\ell}$ (i.e., $\bz(v)$ has independent elements). Define $\mathcal {\widetilde R}$ as $\mathcal {\widetilde R}=\mathcal{R}_0\cup \mathcal{R}_1$ where
$\mathcal{R}_0=\{\bz^r\in\mathcal R: z^r_\ell=1, \ell=1,...,j\}$ and $\mathcal{R}_1 = \{\bz^r\in\mathcal R :\exists \mbox{ one and only one } \ell, s.t., z^r_\ell \neq 1\}.$ Then, for any $0<\epsilon<1$, if $\pi_{\ell, 1}>\frac{q}{q+\sqrt{\epsilon}}$ for all $\ell=1,...,q$, we have $p[\bz(v) \in \mathcal {\widetilde R}]>1-\epsilon$.
\end{theorem}

The proof of the Theorem is relegated to the section 4 of the supplementary material. Based on the above theorem, when $\epsilon \rightarrow 0$, i.e. $p[z_\ell(v)=1] \rightarrow 1$,  we have $p[\bz(v) \in \mathcal {\widetilde R}] \rightarrow 1$. This means that under the given conditions, the probability distribution of the latent state vector  $\bz(v)$ will be restricted to the subspace $ \mathcal {\widetilde R}$. The conditions required in the theorem is well satisfied in fMRI data due to the characteristics of the fMRI neural signals.  Recall that in the MoG source distribution model \eqref{eq:level31}, among the $m$ latent state, we specify $j=1$ as the state corresponding to the background fluctuation. Previous work have established that the fMRI spatial source signals are very sparse across the brain \citep{mckeown1997analysis, daubechies2009independent, lee2011independent}. That is, within a specific BFN(IC, or spatial source signal),  most of the voxels exhibit background fluctuations with only a very small proportion of voxel being activated (or deactivated), which implies $p[z_\ell(v)=1] \rightarrow 1$. Therefore, given the sparsity of the fMRI signals, Theorem \ref{thm1} shows the probability distribution of the latent state vector $\bz(v)$ in our hc-ICA is approximately restricted on the subspace $\mathcal {\widetilde R}$. An implication of this result is that there is little chance for the same voxel to be activated in more than one ICs. Biologically, this means that there is little overlapping in the activated regions across different BFNs, which has been supported by findings in the existing neuroimaging literature.

 Based on this result, we propose a subspace-based approximate EM for our ICA model. The approximate EM follows similar steps as the exact EM. The main difference is that we restrict the condition distribution of the latent state vector $\bz(v)$ to the subspace $\mathcal {\widetilde R}$ in the E-step and M-step. That is, the conditional expectations in the E-step are evaluated with  a subspace-based approximate distribution $\tilde{p}[\bz(v) =z^r| \mathbf {y}(v); \hTh^{(k)}]=p[\bz(v)=z^r | \mathbf {y}(v); \hTh^{(k)}]/\sum_{r \in \mathcal {\widetilde R}}p[\bz(v)=z^r | \mathbf {y}(v); \hTh^{(k)}]$ where $z^r \in \mathcal {\widetilde R}$ (see section 5 of the supplementary material for a detailed treatment). Since the subspace $\mathcal {\widetilde R}$ has a cardinality of $(m-1)q+1$, the approximate EM only requires $\mathcal{O}(mq)$ operations to complete. The concentration of measures to the subspace leads to the simplification in evaluating the conditional expectations in the E-step. For example,
\begin{equation}
\widetilde{E}[\bbs(v)\mid \bby(v);\Theta] = \sum_{\bz(v)\in \mathcal {\widetilde R}}\tilde p[\bz(v) \mid \bby(v);\Theta] E[\bbs(v)\mid \bby(v), \bz(v); \Theta],
\end{equation}
which implies that, instead of summing over $m^q$ latent states in $\mathcal{R}$ , we only need to perform  $(m-1)q+1$ summations across the subspace of $\mathcal {\widetilde R}$. The subspace-based EM also leads to reduction of computation time in the M-step. Specifically, when updating the parameters for the MoG source distribution model, we now use approximate conditional marginal moments. For example, as compared with the exact results, we use the following approximate moment when updating parameters for the Gaussian mixtures,
\begin{equation}
\widetilde{E}[s_{0\ell}(v)\mid z_\ell(v)=j, \bby(v);\Theta]= \frac{\sum_{\bz(v)\in  \mathcal {\widetilde R}^{(\ell, j)}}\tilde p[\bz(v)\mid \bby(v);\Theta]E[s_{0\ell}(v)\mid \bby(v), \bz(v);\Theta]}{\sum_{\bz(v)\in  \mathcal {\widetilde R}^{(\ell, j)}}\tilde p[\bz(v)\mid \bby(v);\Theta]},
\end{equation}
where $\mathcal{\widetilde R}^{(\ell, j)}=\{\bz^r\in \mathcal{\widetilde R}: z^r_\ell=j\}$, whose cardinality equals to $(m-2)q+1$ if $j=1$ and $1$ if $j \neq 1$. Comparing to its exact counterpart, $\mathcal{R}^{(\ell, j)}=\{\bz^r\in \mathcal{ R}: z^r_\ell=j\}$, which has a cardinality of $m^{q-1}$, this can dramatically simplify the updating of $\pi_{\ell,j}, \mu_{\ell,j}$ and $\sigma^2_{\ell,j}$ in the M-step. We summarize the approximate EM algorithm as Algorithm 2.

\begin{algorithm}
   \caption{The Subspace-based Approximate EM Algorithm}
   \label{alg:example}
\begin{algorithmic}
    \STATE {\bfseries Initial values}:  Start with initial values $\hTh^{(0)}$.
   \REPEAT
     \STATE \textbf{E-step:}
        \STATE 1. Determine $\tilde p[\bbs(v) \mid \bby(v);\hTh^{(k)}]$ and its marginals as follows:
        \STATE \quad 1.a Evaluate the multivariate Gaussian $p[\bbs(v) \mid \bby(v), \bz(v); \hTh^{(k)}]$;
        \STATE \quad 1.b Evaluate $\tilde p[\bz(v) \mid \bby(v); \hTh^{(k)}]$ on the subset $\mathcal{\tilde R}$;
        \STATE \quad 1.c $\tilde p[\bbs(v) , \bz(v)\mid \bby(v), \hTh^{(k)}] = p[\bbs(v) \mid \bby(v), \bz(v); \hTh^{(k)}]\times \tilde p[\bz(v) \mid \bby(v); \hTh^{(k)}]$; \\ \quad\quad \ \ $p[\bbs(v)\mid \bby(v), \hTh^{(k)}]=\sum_{\bz(v)\in \mathcal{\tilde R}}\tilde p[\bbs(v) , \bz(v)\mid \bby(v), \hTh^{(k)}] $;
     \STATE 2. Evaluate conditional expectations in $Q(\Theta|\hTh^{(k)})$ with regard to\\ \quad\quad\ \  $\tilde p[\bbs(v) ,\bz(v)|\bby(v);\hTh^{(k)}]$.
     \STATE \textbf{M-step:}
     \STATE Update $\bbeta(v)$, $\bA_i,\pi_{\ell, j}$, $\mu_{\ell, j}$; $\sigma^2_{\ell, j}$ with the modification of replacing the exact conditional moments with their counterparts based on $\tilde p[\bbs(v) \mid \bby(v);\hTh^{(k)}]$.
     \STATE Update $\bm D, \bm E$ with similar modifications of replacing the exact conditional moments with those based on $\tilde p[\bbs(v) \mid \bby(v);\hTh^{(k)}]$.
\UNTIL{$\frac{\|\hTh^{(k+1)}-\hTh^{(k)}\|}{\|\hTh^{(k)}\|} <\epsilon$}
\end{algorithmic}
\end{algorithm}

\subsection{Inference for covariate effects in hc-ICA model}
Typically, statistical inference in maximum likelihood estimation is based on the inverse of the information matrix which is used to estimate the asymptotic variance-covariance matrix of the MLEs. Since Standard EM algorithms only provide parameter estimates, extensions to the EM algorithm have been developed to estimate the information matrix \citep{louis1982finding, meilijson1989fast, meng1991using}. However, these methods are computationally expensive for the proposed hc-ICA model due to the following reasons. First, the dimension of the information matrix for our model is huge due to the large number of parameters. Secondly, the ML estimates, $\hat{\bbeta}(v), v=1,...,V$, are not independent across voxels because they involve the estimates of the same set of parameters such as the mixing matrices. Consequently, the information matrix of the hc-ICA model is ultra-high dimensional and is not sparse, which makes it extremely challenging to invert.

In this section, we present a statistical inference procedure for covariate effects in hc-ICA model. The proposed method is developed based on the connection between the hc-ICA and standard linear models.  Our method aims to provide an efficient approach to estimate the asymptotic standard errors of the covariate effects at each voxel, i.e. $\hat{\bbeta}(v) (v=1,\ldots,V)$, by directly using the output from our EM algorithms. Specifically, we first rewrite the hc-ICA model in a non-hierarchical form by collapsing the two-level models in \eqref{eq:level1} and \eqref{eq:level2} and then multiplying the orthogonal mixing matrix $\bm A_{i}$ on both sides:
\begin{equation}\label{aprox1}\bm A_i' \by_i(v) = \bs_0(v) + \bm X_i\vect\left[\bbeta(v)'\right] + \bm \gamma_i(v) + \bm A_i'\bm  e_i(v),\end{equation}
where $\bm X_i = \bm x_i'\otimes \bm I _q$. \eqref{aprox1} can be re-expressed as follows:
\begin{equation}\label{aprox3} \by_i^*(v)=\bm X_i\vect\left[\bbeta(v)'\right] + \bm \zeta_i(v),\end{equation}
where $\by_i^*(v)=\bm A_i' \by_i(v)-\bs_0(v)$, and $\bm\zeta_i(v)=\bm \gamma_i(v) + \bm A_i'\bm  e_i(v)$ is a multivariate zero-mean Gaussian noise term. The model in \eqref{aprox3} can be viewed as a general multivariate linear model at each voxel. The major distinction of \eqref{aprox3} from the standard linear model is that the dependent variable $\by^*(v)$ not only depends on the observed data $\by(v)$ but also involves unknown parameters $\bm A_i$ and latent variables $\bs_0(v)$. Given the similarity between hc-ICA and the standard linear model, we propose a variance estimator for $\vect\left[\hat{\bbeta}(v)'\right]$ following the linear model theory.

Note that, for a standard linear model, the asymptotic variance for $\vect\left[\hat{\bbeta}(v)'\right]$ can be obtained by:
\begin{equation}\label{avar}
\mbox{Var}\left\{\vect\left[\hat{\bbeta}(v)'\right]\right\}=\frac{1}{N}\left(\sum_{i=1}^N \bm X_i' \bm W(v)^{-1}\bm X_i\right)^{-1},\end{equation}
where $\bm W(v)$ is the variance of the Gaussian noise in the linear model. Then, the variance of $\vect\left[\hat{\bbeta}(v)'\right]$ can be estimated by plugging in an estimator for $\bm W(v)$ in (\ref{avar}).   Following this result, we consider a variance estimator for $\vect\left[\hat{\bbeta}(v)'\right]$ based on (\ref{avar}) by plugging in the empirical variance estimator $\widehat{\bm W}(v) = \frac{1}{N}\sum_{i=1}^N\left(\by_i^{*}(v)-\bm X_i\vect\left[\hat{\bbeta}(v)'\right]\right)^{\otimes 2}$ \citep{seber2012linear}. Because the dependent variable $\by^*(v)$ in \eqref{aprox3} is not directly observable, we estimate $\by_i^{*}(v)$ using the ML estimates from our EM algorithm as $\widehat{\by_i^{*}}(v)=\widehat{\bm A}_i' \by_i(v)-\widehat{\bs_0}(v)$, where $\widehat{\bs_0}(v)=E[\bs_0(v)|\bby(v), \hTh]$. That is, we modify the empirical variance estimator $\widehat{\bm W}(v)$ as follows:
\begin{equation}\label{hatV}
\widetilde{\bm W}(v) = \frac{1}{N}\sum_{i=1}^N
 \left(\widehat{\bm A}_i' \by_i(v)-E[\bs_0(v)|\bby(v), \hTh]-\bm X_i\vect\left[\hat{\bbeta}(v)'\right]\right)^{\otimes 2}.
\end{equation}
Thus, our final variance estimator is $\widehat{\mbox{Var}}\left\{\vect\left[\hat{\bbeta}(v)'\right]\right\}=\frac{1}{N}\left(\sum_{i=1}^N \bm X_i' \widetilde{\bm W}(v)^{-1}\bm X_i\right)^{-1}$. 

We then can perform hypothesis testing on the covariate effects at each voxel by calculating the Z-statistics based on the proposed variance estimator and determine the corresponding $p$-values. Our method can test whether a certain covariate has significant effects on each of the BFNs at the voxel level. Based on the parametric Z-statistic maps, one can also apply standard multiple testing methods to control the family wise error rate (FWER) or the false discovery rate (FDR) in testing the covariate effects within a BFN.

\section[]{Simulation Study}\label{sec3}
We conducted three sets of simulation studies to 1) evaluate the performance of the proposed hc-ICA model as compared with the existing TC-GICA model, 2) to compare the accuracy of the subspace-based approximate EM algorithm vs. the exact EM, 3) and to evaluate the performance of the proposed inference method for covariate effects based on hc-ICA.

\subsection{Simulation Study I: performance of the hc-ICA vs. TC-GICA}
In the first simulation study, we evaluated the performance of the proposed hc-ICA model as compared with a popular TC-GICA two-stage method: the dual regression ICA \citep{beckmann2009group}. We simulated fMRI data from three underlying source signals, i.e., $q=3$, and considered three sample sizes with the number of subjects of $N=10, 20, 40$. For each source, we generated a 3D spatial map with the dimension of $25\times 25\times 4$ (Figure \ref{fig1}(A)). For spatial source signals, we first generated population-level spatial maps, i.e. $\{\bs_0(v)\}$, as the true source signals plus Gaussian random noise of variance 0.5. We then generated two covariates for each subject with one being categorical ($x_{1}\iid \mbox{Bernoulli}(0.5)$) and the other being continuous ($x_{2}\iid \mbox{Uniform}(-1,1)$). The covariate effects maps, i.e.$\{\bbeta(v)\}$, are presented in Figure \ref{fig1}(B1)-(B2) where the covariate effect parameters at each voxel took values from $\{0,1.5,1.8,2.5,3.0\}$. Additionally, we generated Gaussian subject-specific random effects, i.e. $\bm \gamma_i(v)$, and considered three levels of between-subject variability: low ($\bm D = \mbox{diag}(0.1, 0.3, 0.5)$), medium ($\bm D = \mbox{diag}(1.0, 1.2, 1.4)$) and high ($\bm D = \mbox{diag}(1.8, 2.0, 2.5)$). The subject-specific spatial source signals were then simulated as the linear combination of the population-level signals, covariate effects and subject-specific random effects. For temporal responses, each source signal had a time series of length of $T=200$ that was generated based on time courses from real fMRI data and hence represented realistic fMRI temporal dynamics. We generated subject-specific time sources that had similar frequency features but different phase patterns \citep{guo2011general}, which represented temporal dynamics in resting-state fMRI signals. After simulating the spatial maps and time courses for the source signals, Gaussian background noise with a standard deviation of 1 were added to the source signals to generate observed fMRI data.

Based on the simulated data, we compared the performance of the proposed hc-ICA model and the dual-regression ICA. Following previous work \citep{beckmann2005tensorial, guo2008unified, guo2011general}, we evaluated the performance of each method based on the correlations between the true and estimated signals in both temporal and spatial domains. Furthermore, to compare the performance in estimating the covariate effects, we report the mean square errors (MSEs) of $\hat{\bbeta}(v)$ defined based on  $\left\|\hat{\bbeta}(v)-\bbeta(v)\right\|_{\mathcal{F}}^2$ averaged across simulation runs. Here $\|\cdot\|_{\mathcal{F}}$ is the Frobenius norm for matrix. Since ICA recovery is permutation invariant, each estimated IC was matched with the original source with which it had the highest spatial correlation. We present the simulation results in Table \ref{tbl1}. The results show that the hc-ICA provides more accurate estimates for the source signals on both the population- and subject-level and also has smaller mean square errors in estimating the covariate effects. We also display the estimated population-level IC maps and the covariate effects maps from both methods in Figure \ref{fig1}. The figure shows that the hc-ICA showed much better performance in correctly detecting the true distributed patterns and covariate effects for each IC. In comparison, the estimates of the population-level IC maps from the dual-regression were contaminated by the covariate effects. Furthermore, the estimated covariate effects maps based on the dual regression were much noisier and demonstrated some mismatches across the ICs due to the low correlation between the true and estimated spatial IC maps.

{\small
 \begin{table}[h!]
\begin{center}
\caption{Simulation results for comparing our hc-ICA method against the dual-regression ICA based on 100 runs. Values presented are mean and standard deviation of correlations between the true and estimated: subject-specific spatial maps, population-level spatial maps and subject-specific time courses. The mean and standard deviation of the MSE of the covariate estimates are also provided.}\label{tbl1}
\begin{tabular}{lccccc}
\toprule
\hline
 Btw-subj         &\multicolumn{2}{c}{Population-level spatial maps} &   &\multicolumn{2}{c}{Subject-specific spatial maps}    \\
 Var              & \multicolumn{2}{c}{Corr.(SD)}                     &   &\multicolumn{2}{c}{Corr.(SD)}                      \\
                   \cline{2-3}     \cline{5-6}
                  &    hc-ICA       & Dual.Reg.                     &   &     hc-ICA      & Dual.Reg.     \\
\hline
Low & & & & & \\
         N=10    &  0.982 (0.003)   & 0.956 (0.018)   &   &  0.984 (0.004)   &  0.945 (0.023)      \\
         N=20    &  0.990 (0.002)   & 0.968 (0.014)   &   &  0.996 (0.002)   &  0.949 (0.008)      \\
         N=40    &  0.992 (0.002)   & 0.976 (0.005)   &   &  0.996 (0.001)   &  0.956 (0.002)      \\  \\
Medium & & & & &  \\
         N=10    & 0.942 (0.017)    & 0.914 (0.048)  &   &  0.943 (0.011)    &  0.882 (0.030)       \\
         N=20    & 0.954 (0.002)    & 0.938 (0.034)  &   &  0.959 (0.004)    &  0.890 (0.016)       \\
         N=40    & 0.961 (0.002)    & 0.949 (0.020)  &   &  0.968 (0.003)    &  0.893 (0.009)     \\  \\
High & & & & &  \\
         N=10    & 0.833 (0.146)    & 0.740 (0.164)  &   &  0.894 (0.108)    &  0.689 (0.303)       \\
         N=20    & 0.850 (0.129)    & 0.795 (0.143)  &   &  0.909 (0.084)    &  0.695 (0.281)       \\
         N=40    & 0.871 (0.055)    & 0.809 (0.102)  &   &  0.928 (0.035)    &  0.705 (0.259)     \\
     \midrule
Btw-subj         &    \multicolumn{2}{c}{Subject-specific time courses}&  & \multicolumn{2}{c}{Covariate Effects}      \\
Var.             & \multicolumn{2}{c}{Corr.(SD)}                        &  & \multicolumn{2}{c}{MSE(SD)}  \\
                 \cline{2-3}         \cline{5-6}
                 &    hc-ICA       & Dual.Reg.                      &   &   hc-ICA      & Dual.Reg.     \\
                  \hline
Low & & & & & \\
         N=10    & 0.998 (0.001)   &  0.987 (0.010)      &  & 0.048 (0.019) &  0.154 (0.055)\\
         N=20    & 0.998 (0.001)   &  0.995 (0.004)      &  & 0.021 (0.003) &  0.127 (0.044)\\
         N=40    & 0.998 (0.001)   &  0.994 (0.004)      &  & 0.012 (0.001) &  0.111 (0.030)\\   \\
Medium & & & & &  \\
         N=10    & 0.993 (0.010)   &  0.970 (0.028)      &  & 0.273 (0.088) &  0.485 (0.151)  \\
         N=20    & 0.998 (0.003)   &  0.976 (0.016)      &  & 0.117 (0.015) &  0.285 (0.076) \\
         N=40    & 0.998 (0.002)   &  0.991 (0.008)      &  & 0.064 (0.005) &  0.187 (0.041) \\\\
High & & & & &  \\
         N=10    & 0.948 (0.021)   &  0.903 (0.045)      &  & 0.387 (0.157) &  0.783 (0.325)  \\
         N=20    & 0.978 (0.018)   &  0.925 (0.029)      &  & 0.224 (0.075) &  0.532 (0.271) \\
         N=40    & 0.990 (0.015)   &  0.934 (0.022)      &  & 0.131 (0.056) &  0.389 (0.198) \\
\hline\bottomrule
\end{tabular}
\end{center}
\end{table}
         }

\begin{figure}[p]
\begin{center}$
\small
\begin{array}{ccc}
\hline
\multicolumn{3}{c}{\mbox{\small{ (A) Population-level IC maps }} }\\
 \includegraphics[scale=0.2]{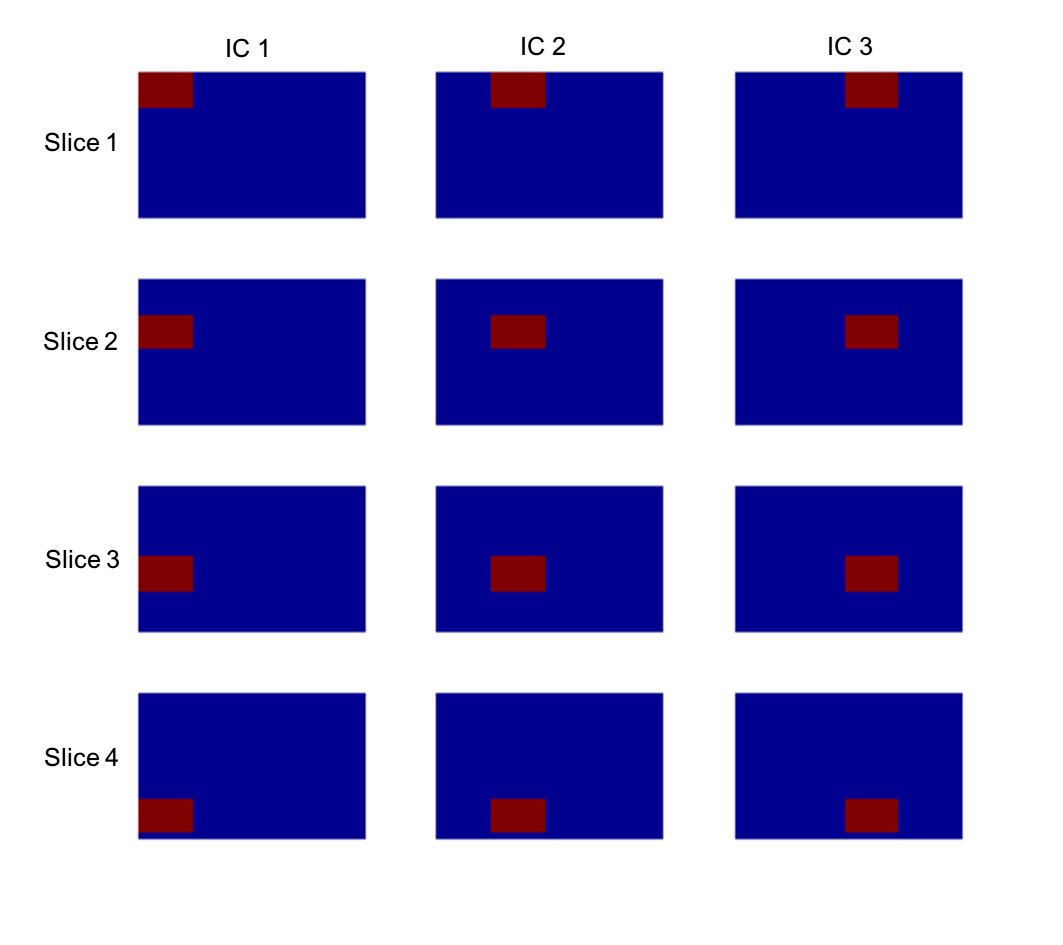}
 &\includegraphics[scale=0.2]{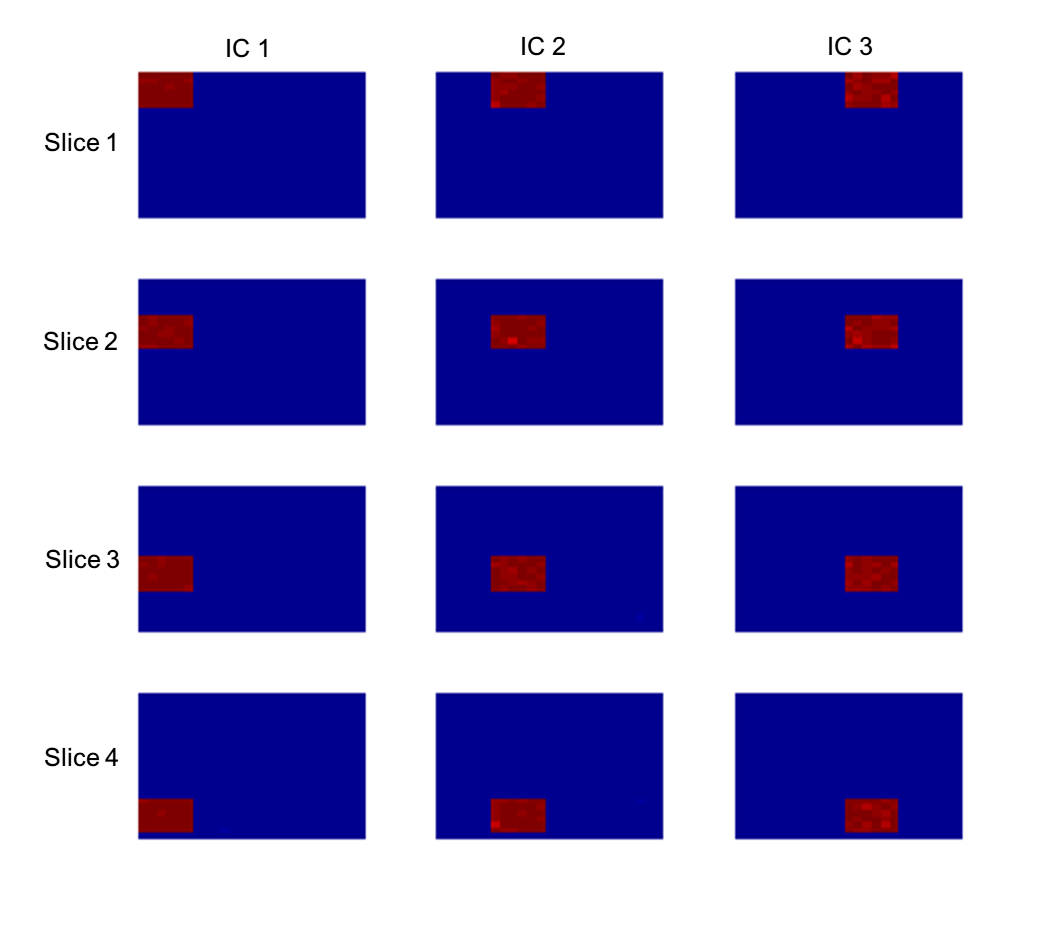}
 &\includegraphics[scale=0.2]{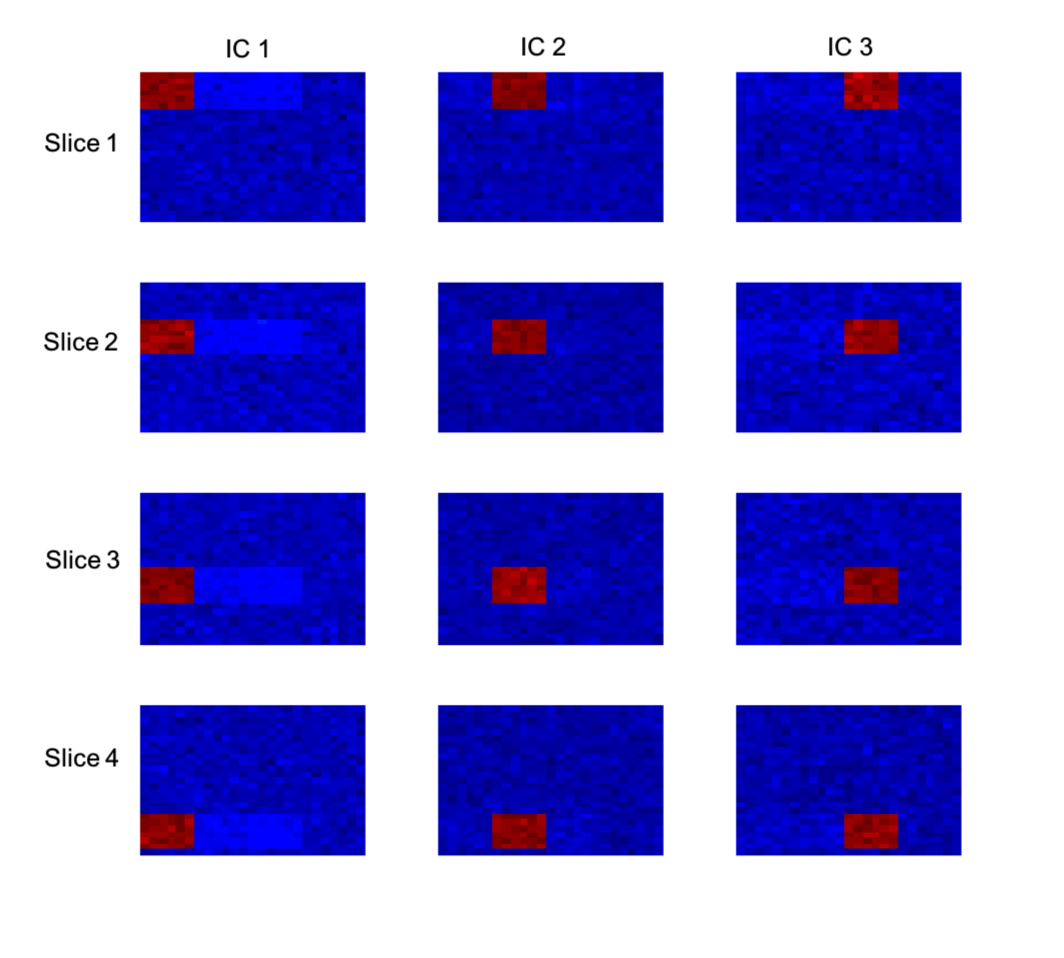} \\
 \mbox{Truth} &\mbox{hc-ICA} &\mbox{Dual.Reg.} \\
  \multicolumn{3}{c}{\includegraphics[scale=0.3]{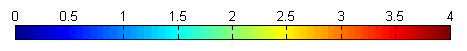}} \\
  \hline\\
 \multicolumn{3}{c}{\mbox{(B1) Covariate effects of the binary covariate, } \bm x_1}\\
 \includegraphics[scale=0.2]{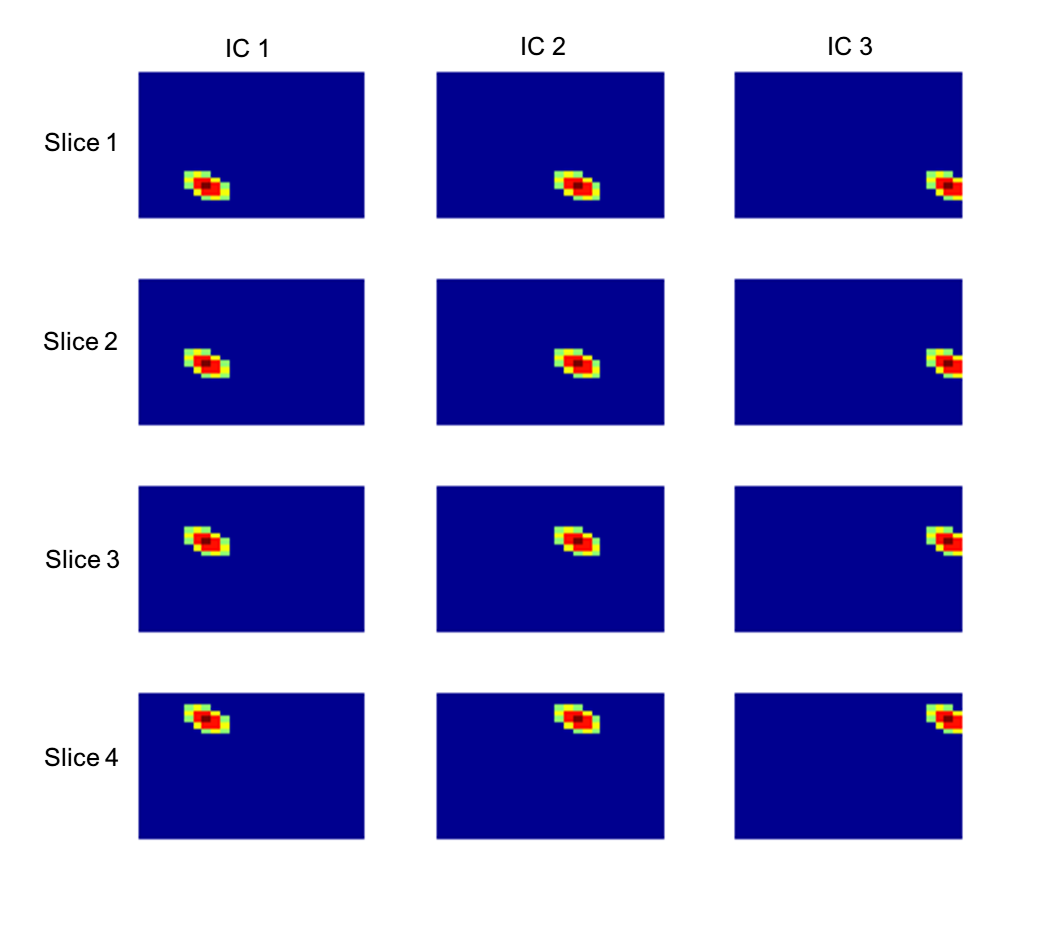}
 &\includegraphics[scale=0.2]{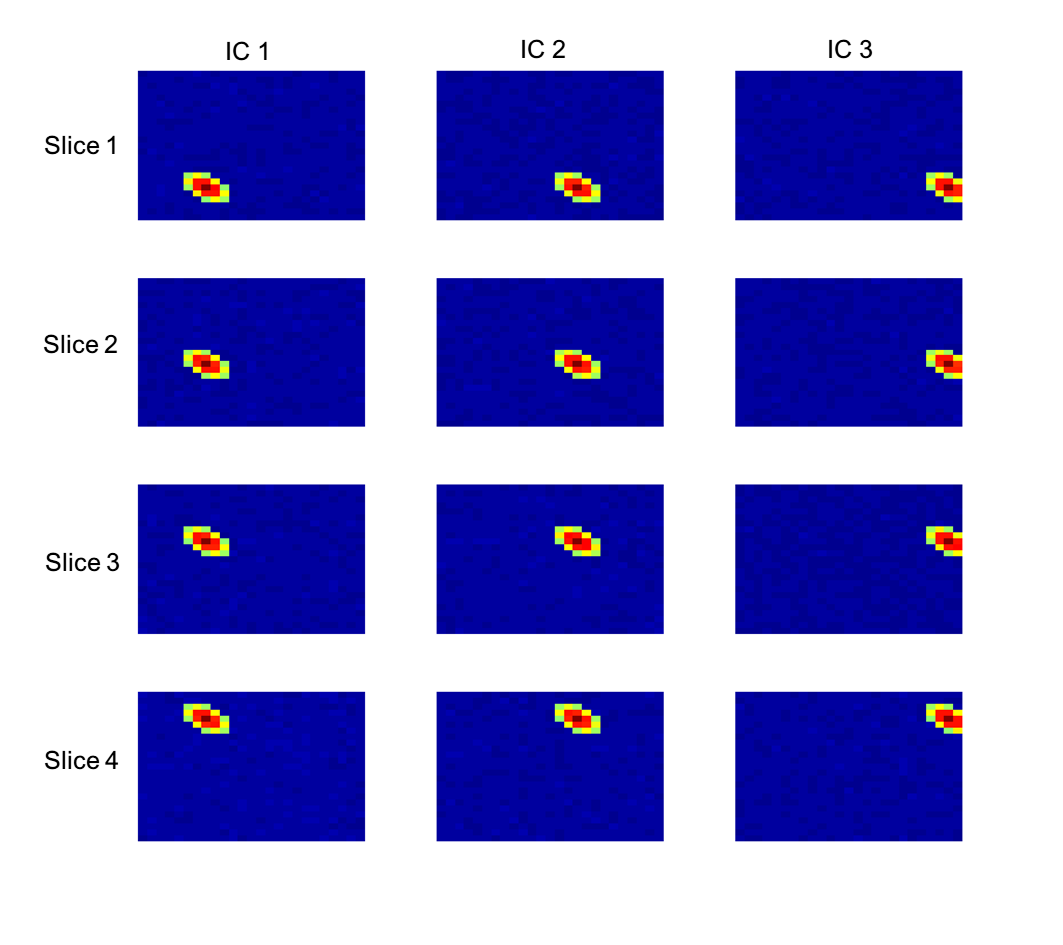}
 &\includegraphics[scale=0.2]{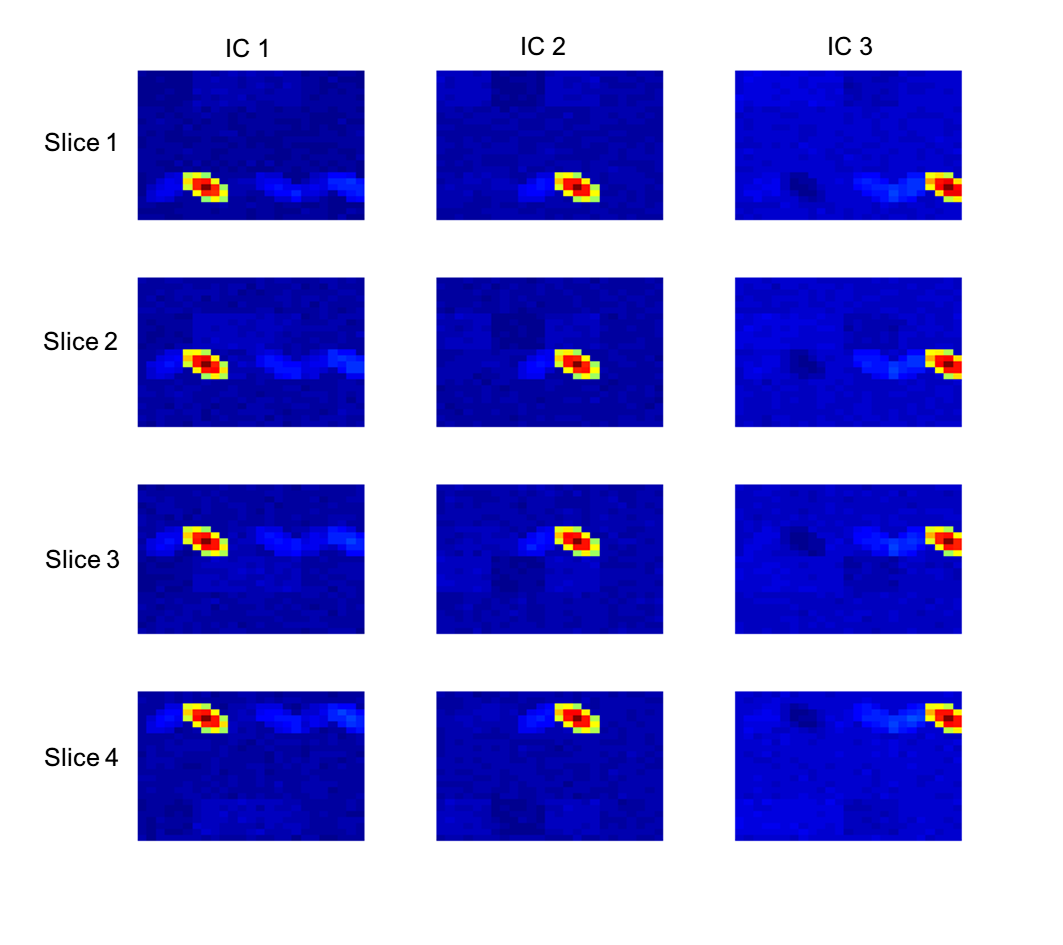} \\
\mbox{ Truth} &\mbox{ hc-ICA} &\mbox{ Dual.Reg.}\\
  \multicolumn{3}{c}{\includegraphics[scale=0.3]{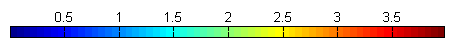}}\\
  \hline\\
 \multicolumn{3}{c}{\mbox{(B2) Covariate effects of the continuous covaraite, } \bm x_2}\\
 \includegraphics[scale=0.2]{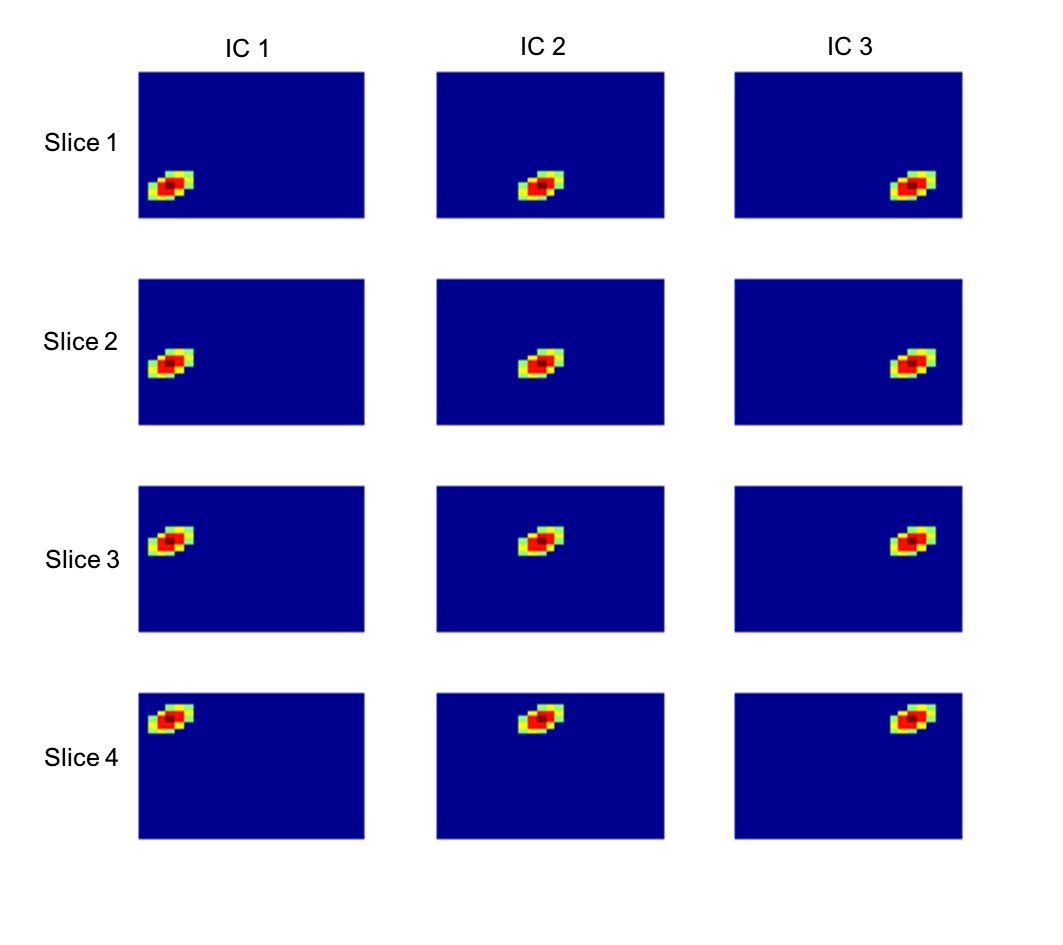}
&\includegraphics[scale=0.2]{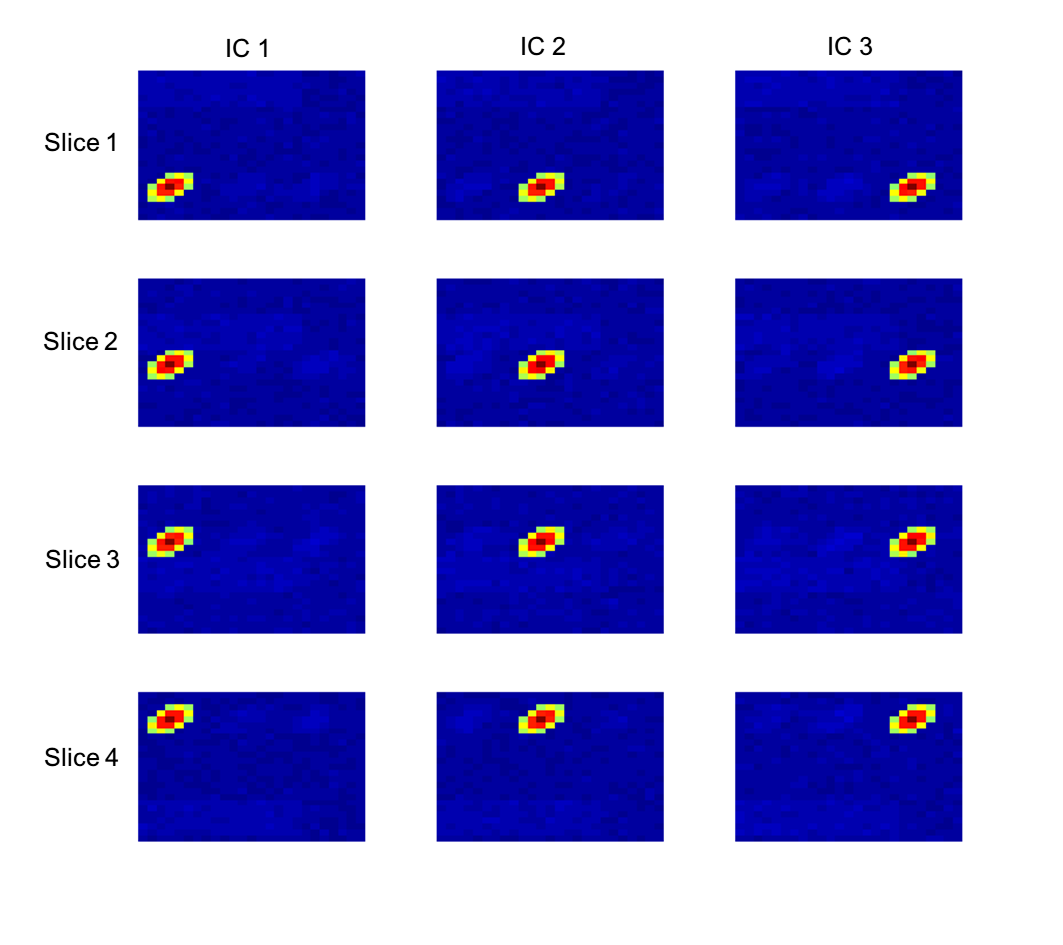}
 &\includegraphics[scale=0.2]{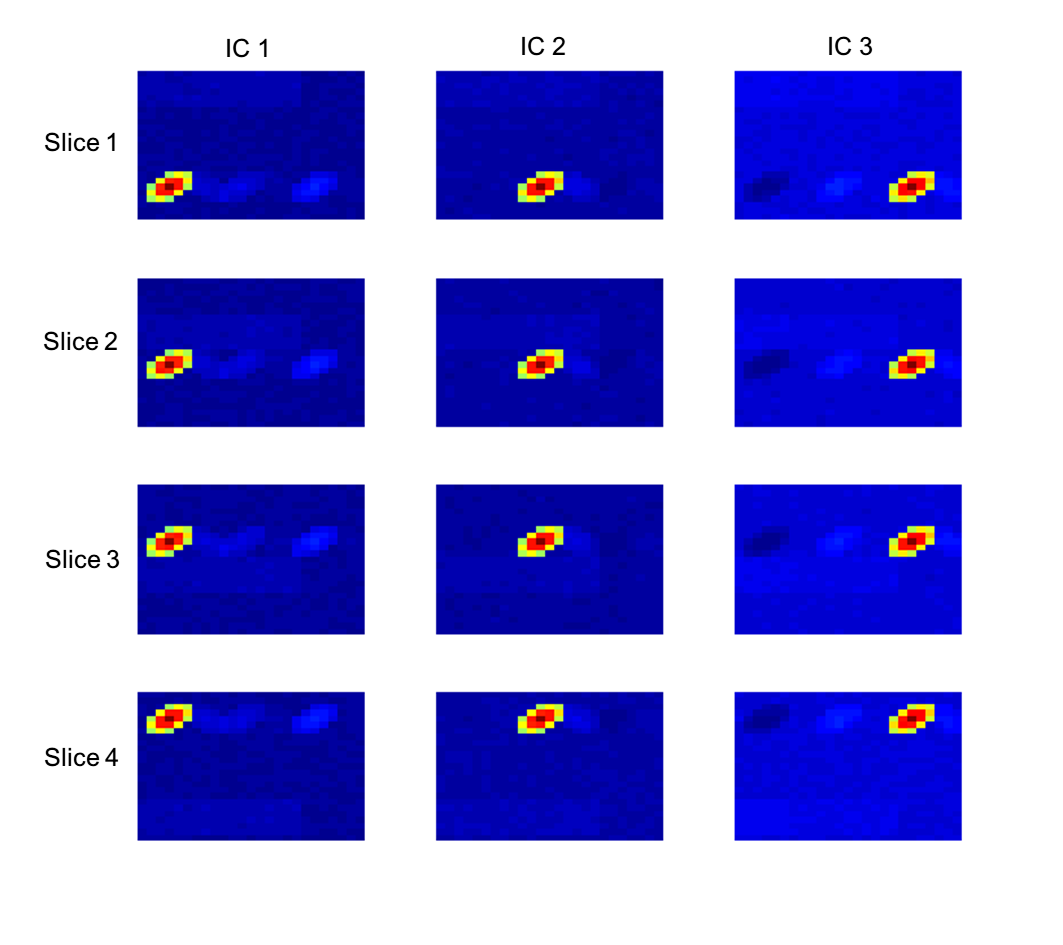} \\
\mbox{ Truth} &\mbox{ hc-ICA} &\mbox{ Dual.Reg.}\\
 \multicolumn{3}{c}{\includegraphics[scale=0.3]{colorbar1.png}}\\
 \hline
\end{array}$
\caption{\small{Comparison between our method and the dual-regression ICA: truth, estimates from our model, estimates from the dual regression (N=10, between-subject variabilities are medium) based on 100 runs. All the images displayed are averaged across the 100 Monte Carlo data sets. Population-level spatial maps are shown in Figure 1(A). The results of the dual-regression ICA are contaminated by the covariate effects. The results from our method are more accurate. Covariate effect estimates are shown in Figure 1(B1) and Figure 1(B2) respectively. The results of the dual-regression show clear mismatching while our method provide accurate estimates.}}\label{fig1}
\end{center}
\end{figure}

\subsection{Simulation Study II: performance of the approximate EM vs. the exact EM}
In the second simulation study, we compared the performance of the exact EM algorithm with the approximate EM for the hc-ICA model. We simulated fMRI data for ten subjects and considered three model sizes with the number of source signals of $q=3, 6, 10$. The fMRI data were generated using methods similar to that in Simulation Study I. We then fit the proposed hc-ICA model using both the exact EM and the approximate EM. Results from Table \ref{tbl2} show that the accuracy of the subspace-based EM was fairly comparable to that of the exact EM in both the spatial and temporal domains and on both population- and subject-level. The convergence rates across simulations were almost the same between the two algorithms. The major advantage of the subspace-based EM is that it was much faster than the exact EM. This advantage became more significant with the increase of the number of source signals. For $q = 10$, the subspace-based EM only used about $2\%$ computation time of the exact EM.

{\small
\begin{table}[h!]
\begin{center}
\caption{Simulation results for comparing the subspace-based approximate EM and the  exact EM based on 50 runs. Mean and standard deviation of correlations between the true and estimated spatial maps and time courses are presented. The mean and standard deviation of the MSE of the covariate estimates are also provided. }\label{tbl2}
\begin{tabular}{lccccc}
\toprule
\hline
     &\multicolumn{2}{c}{Population-level spatial maps}&  & \multicolumn{2}{c}{Subject-specific spatial maps}       \\
 & \multicolumn{2}{c}{Corr(SD)}                    &  & \multicolumn{2}{c}{Corr(SD)}                   \\
                   \cline{2-3}     \cline{5-6}
 \# of IC                   &     Exact EM       & Approx. EM                      &   &    Exact EM      & Approx. EM     \\
\hline

         q=3    & 0.981(0.003)   &  0.981(0.001)     &  & 0.986(0.004) &  0.981(0.002)\\
         q=6    & 0.980(0.006)   &  0.980(0.006)     &  & 0.985(0.012) &  0.981(0.011)\\
         q=10   & 0.969(0.022)   &  0.963(0.020)     &  & 0.972(0.027) &  0.970(0.022)\\   \midrule
&    \multicolumn{2}{c}{Subject-specific time courses}&  & \multicolumn{2}{c}{Covariate Effects}      \\
    & \multicolumn{2}{c}{Corr(SD)}                    &  & \multicolumn{2}{c}{MSE(SD)}                   \\
                   \cline{2-3}     \cline{5-6}
 \# of IC                   &     Exact EM       & Approx. EM                      &   &    Exact EM      & Approx. EM     \\
\hline

         q=3    & 0.998(0.001)   &  0.998(0.000)     &  & 0.048(0.020) &  0.048(0.019)\\
         q=6    & 0.997(0.003)   &  0.995(0.002)     &  & 0.069(0.024) &  0.070(0.022)\\
         q=10   & 0.992(0.016)   &  0.992(0.009)     &  & 0.105(0.033) &  0.112(0.028)\\   \midrule

           & \multicolumn{2}{c}{Time in miniute}&  & \multicolumn{2}{c}{Proportions of Convergence}      \\

                 \cline{2-3}         \cline{5-6}
\# of IC                  &     Exact EM       & Approx. EM                      &   &    Exact EM      & Approx. EM     \\
                  \hline
         q=3    & 9.91  & 5.22    &  & 100\% &  100\%\\
         q=6    & 71.05 & 9.09    &  & 100\% &  100\%\\
         q=10   & 860.10& 19.02   &  & 96\%  &  96\%\\   \hline\bottomrule
\end{tabular}
\end{center}
\end{table}
}
\subsection{Simulation Study III: performance of the proposed inference procedures for covariate effects}
We examine the performance of our inference procedures for $\hat{\bbeta}(v)$ in the third simulation study. We simulated fMRI datasets with two sources signals and considered sample sizes of $N=20, 40, 80$. We generated two covariates in the same manner as in Simulation Study I. To facilitate computation, we generated images with the dimension of $20\times 20$. The variance of between-subject random variabilities was set as $0.25$ for both spatial source signals and the variance of the noise terms added to generate the observed signals was $0.4$. We applied our hc-ICA method and the dual-regression ICA for the simulated datasets and tested for the covariate effects using both methods. The hypotheses were $H_0: \beta_{k\ell}(v) =0$ versus $H_1: \beta_{k\ell}(v) \neq 0$ at each voxel. Specifically, for hc-ICA, hypothesis tests were conducted for $\bbeta(v)$  using the test proposed in section 2.5. In comparison, the dual-regression method tested covariate effects by performing post-ICA regressions of the estimated subject-specific IC maps on the permuted covariates. We estimated the Type-I error rate with the empirical probabilities of not rejecting $H_0$ at voxels such that $\beta_{k\ell}(v)=0$. We report the average of the Type-I error rates at various significance levels, i.e $\alpha$, in Table \ref{tbl3}. We also estimated the power of the tests with the empirical probabilities of rejecting $H_0$ at voxels with non-zero values for the covariate effects parameters, i.e. $\beta_{k\ell}(v) \in \{0,1.5,1.8,2.5,3.0\}$. Results from Table \ref{tbl3} show that our inference method demonstrated lower type-I error rate as well as higher statistical power as compared with the dual-regression ICA. It implies that the proposed inference method based on hc-ICA model provides more accurate inferences for covariate effects on BFNs than the TC-GICA based dual-regression method.

{\small
 \begin{table}[h!]
\begin{center}
\caption{Simulation results for the inference of $\bbeta(v)$ based on 1000 runs. Type-I errors are averaged across all voxels with $\beta_{k\ell}(v)=0$; powers are averaged across voxels having the same values of $\beta_{k\ell }(v)\neq 0$.}\label{tbl3}
\begin{tabular}{cccccccccccc}
\toprule\hline

            &\multicolumn{2}{c}{N=20} && \multicolumn{2}{c}{N=40}  &&  \multicolumn{2}{c}{N=80}  \\\hline

\multicolumn{3}{l}{\emph{Type-I error analysis:}}&&  && \\
size      &hc-ICA &Dual.Reg.   &&hc-ICA &Dual.Reg    &&hc-ICA &Dual.Reg \\ \hline
$0.01$   &$0.014$  &$0.029$   &&0.012   &0.025      &&0.012    &0.018\\
$0.05$   &$0.062$  &$0.084$   &&0.056   &0.076      &&0.055    &0.062\\
$0.10$   &$0.129$  &$0.205$   &&0.118   &0.190      &&0.112    &0.149\\
$0.50$   &$0.522$  &$0.580$   &&0.516   &0.565      &&0.514    &0.557\\
$0.80$   &$0.835$  &$0.872$   &&0.820   &0.856      &&0.810    &0.840\\ \midrule
\multicolumn{3}{l}{\emph{Power analysis (test size: $0.05$):}}&&  && \\
$\bbeta(v) $&hc-ICA &Dual.Reg.     &&hc-ICA &Dual.Reg        &&hc-ICA &Dual.Reg \\
\hline
$1.5$  &0.144  &0.130   &&0.256   &0.203  &&0.404 &0.284\\
$1.8$  &0.268  &0.224   &&0.474   &0.390  &&0.812 &0.548\\
$2.5$  &0.589  &0.475   &&0.862   &0.705  &&0.963 &0.839\\
$3.0$  &0.907  &0.845   &&1.000   &0.922  &&1.000 &1.000\\\hline\bottomrule

\end{tabular}
\end{center}
\end{table}
         }

\section[]{fMRI study of brain functional networks among Zen meditators}\label{sec4}
In recent years, there has been strong interest in neuroscience community to investigate whether Zen meditation practices can potentially contribute to the phenomenological and epistemological aspects of cognitive sciences \citep{pagnoni2007age, lutz2008attention, holzel2011mindfulness, gard2014potential}. We apply our methods to an fMRI study of the effect of Zen meditation. In this study, twelve Zen meditators with more than 3 years of daily practice were recruited along with twelve control subjects who have never practiced meditation. The groups were matched for gender (MEDT, 10 Male; CTRL, 9 Male), age (mean$\pm$SD: MEDT, 37.3$\pm$7.2 years; CTRL, 35.3$\pm$5.9 years; 2-tailed, 2-sample t-test: $p=0.45$), and education level (mean$\pm$SD: MEDT, 17.8$\pm$2.5 years; CTRL, 17.6$\pm$1.6 years; $p=0.85$). All participants were native English speakers and right-handed, except for one ambidextrous meditator.

The hypothesis in this study is that meditators would present different spatio-temporal brain functional activities compared with the control when exposed to automatic conceptual processing tasks. The fMRI study adapted a simple lexical decision paradigm employing semantic and nonsemantic stimuli following \citet{binder2003neural}. In the experiment, 50 words and 50 phonologically and orthographically matched nonword items were presented visually on a screen in pseudo-random temporal order. The subjects were asked to respond whether the displayed item was ``a real English word" via a button-box with their left hand (index finger = "yes", middle finger = "no"). Subjects were instructed to use the awareness of their breathing throughout the session as a reference point to monitor and counteract attentional lapses. The experimental task can be thus conceived as having a dual-layer structure: an ongoing meditative baseline condition and a phasic perturbation of this baseline by semantic and nonsemantic stimuli. For each subject, a T1-weighted high-resolution anatomical image (MPRAGE, 176 sagittal slices, voxel size: $1\times 1\times 1$ mm) and a series of functional images (echo-planar, 520 scans, TR=2.35s, TE=30, voxel size: $3\times 3\times 3$ mm) were acquired on a 3.0 Tesla Siemens Magnetom Trio scanner. Each of the $520$ fMRI scans contained $53\times 63\times 46$ voxels. The fMRI images were corrected for slice acquisition time and subject movements, registered to the mean of the corrected functional images and spatially normalized to the MNI standard brain space using SPM5 (\url{http://fil.ion.ucl.ac.uk/spm/software/spm5}). The computed MNI-normalization parameters were then applied to smooth the functional images with an 8 mm isotropic Gaussian kernel.

Prior to hc-ICA, we performed preprocessing steps including centering, dimension reduction and whitening as described in section 2.1 where the number of ICs was chosen to be 14 based on Laplace approximation \citep{minka2000automatic}. The preprocessed fMRI data from the subjects in the meditation and control groups were then decomposed using the proposed hc-ICA model. Given that the subjects in the meditation and control groups were matched on relevant demographic variables, we included the group indicator as the covariate in the second-level model of hc-ICA. The hc-ICA model was estimated using the subspace-based EM algorithm implemented by in-house MATLAB programs developed by the authors, which will be made available at the authors' website. The computation time was around 4 hours on a Sun GridEngine cluster with 32 nodes. The initial values for the EM algorithm were specified using results from existing group ICA package. To test the robustness of EM results under different initial settings, we ran the algorithm multiple times with different sets of initial values by introducing noises to original initial values, randomly shuffling the orders of the values and randomly changing the signs of the parameters. The results from our EM remained quite stable under the different initial settings (see the last section of the supplementary material for more details).

\subsection{Brain functional networks for the task-related fMRI}
Among the extracted ICs, two functional networks were closely related to the neural systems involved in the experimental task and the meditation practice. The first network included the supplementary motor area (SMA), the hand region of the right sensorimotor cortex (HRSC, contralateral to the left hand pressing the button) and the visual cortex (VC). We labeled this network as the task-related network (TRN) because the functions of the activated regions were clearly associated with the experimental tasks in this study and the temporal dynamics of this network had the highest correlation with the task time series. The second functional network of interest included the posterior cingulate cortex (PCC), the medial prefrontal cortex (MPC), the left lateral parietal cortex (LLPC) and the hippocampus (HIP). This network is well-known as the ``default mode network (DMN)" \citep{raichle2001default} which features increased metabolism at resting states and decreased activities during active tasks.

\begin{figure}[h]
	\begin{center}
		\small
		\includegraphics[scale=0.42]{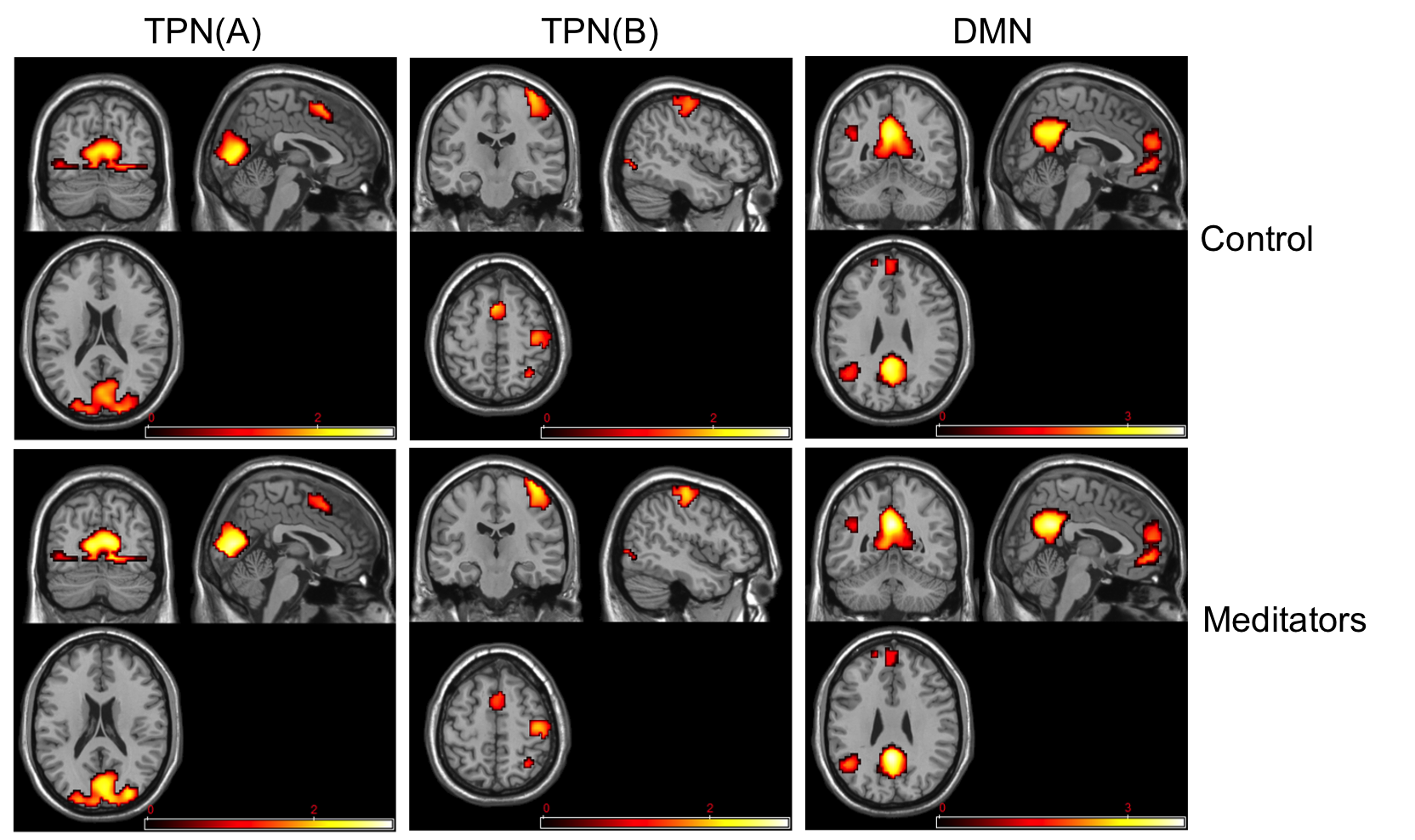}
		\caption{Subgroup brain functional network maps for the task-related network (TRN) and default mode network (DMN) for the control and meditators (threholded by conditional probability of activation larger than $0.95$). Here and below, TPN(A) demonstrates VC and SMA; TPN(B) demonstrates HRSC and SMA; DMN demonstrates all the related structures: PCC, MPC, LLPC and HIP. }\label{fig2}
	\end{center}
\end{figure}

Compared with previous findings in \cite{guo2013hierarchical}, one major benefit of hc-ICA is that it can provide model-based subgroup estimates for the brain networks. Figure \ref{fig2} shows the spatial maps of these two networks for mediators and control subjects. The activated brain regions in each network included voxels with an estimated conditional probability of activation exceeding $0.95$ (see section 6 in the supplementary material for more details). Figure \ref{fig2} indicates meditators generally had stronger signals in these networks as compared with the control. In the following part, we present the results from formal statistical tests of the group effects on the functional networks based on hc-ICA.

\subsection{The effects of Zen meditation on brain functional networks}
\begin{figure}[!h]
	\begin{center}
		$\begin{array}{c}
		\small
		\includegraphics[scale=0.28]{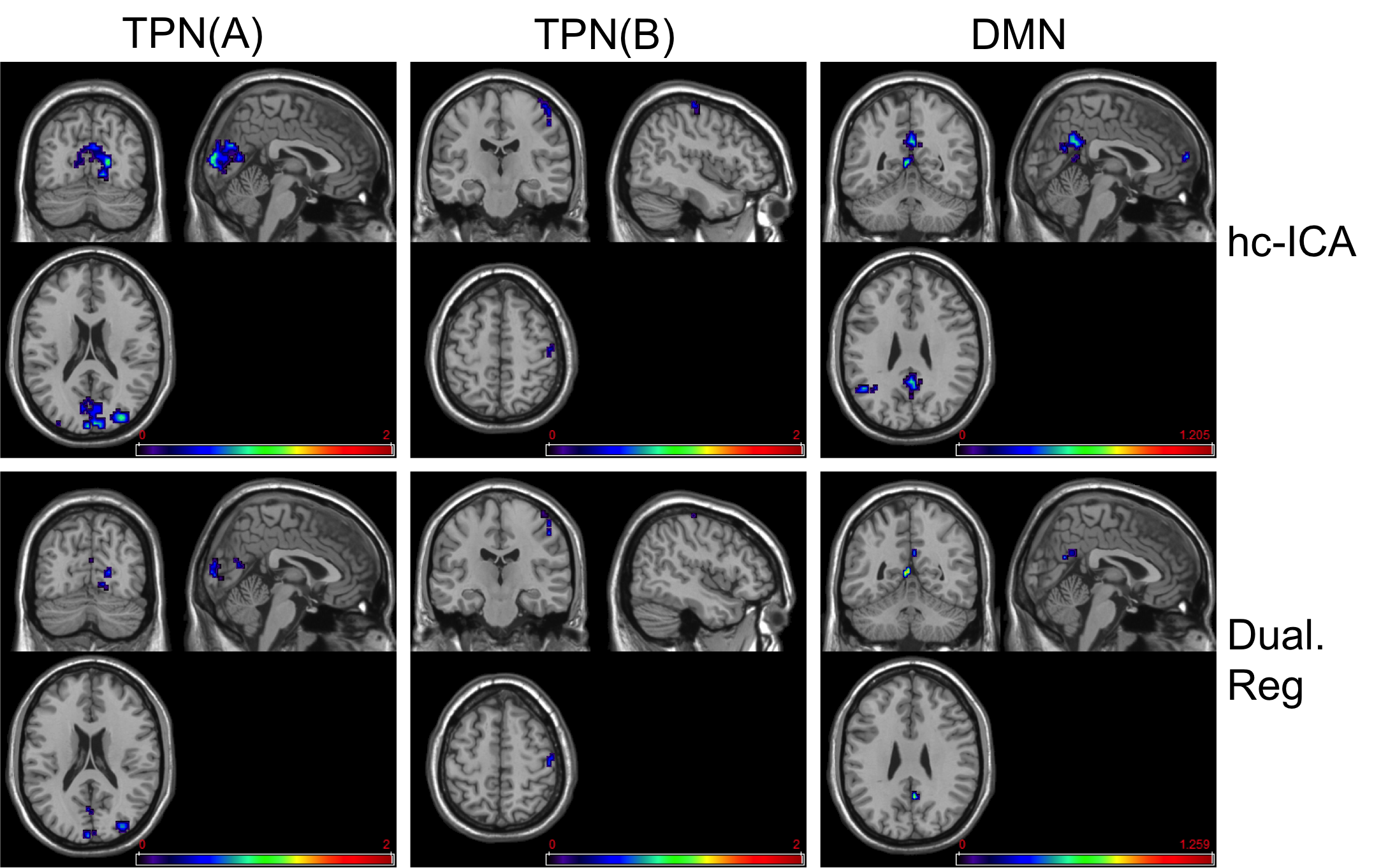}\\
		\mbox{estimated between-group differences}\\
		\hline\\
		\includegraphics[scale=0.28]{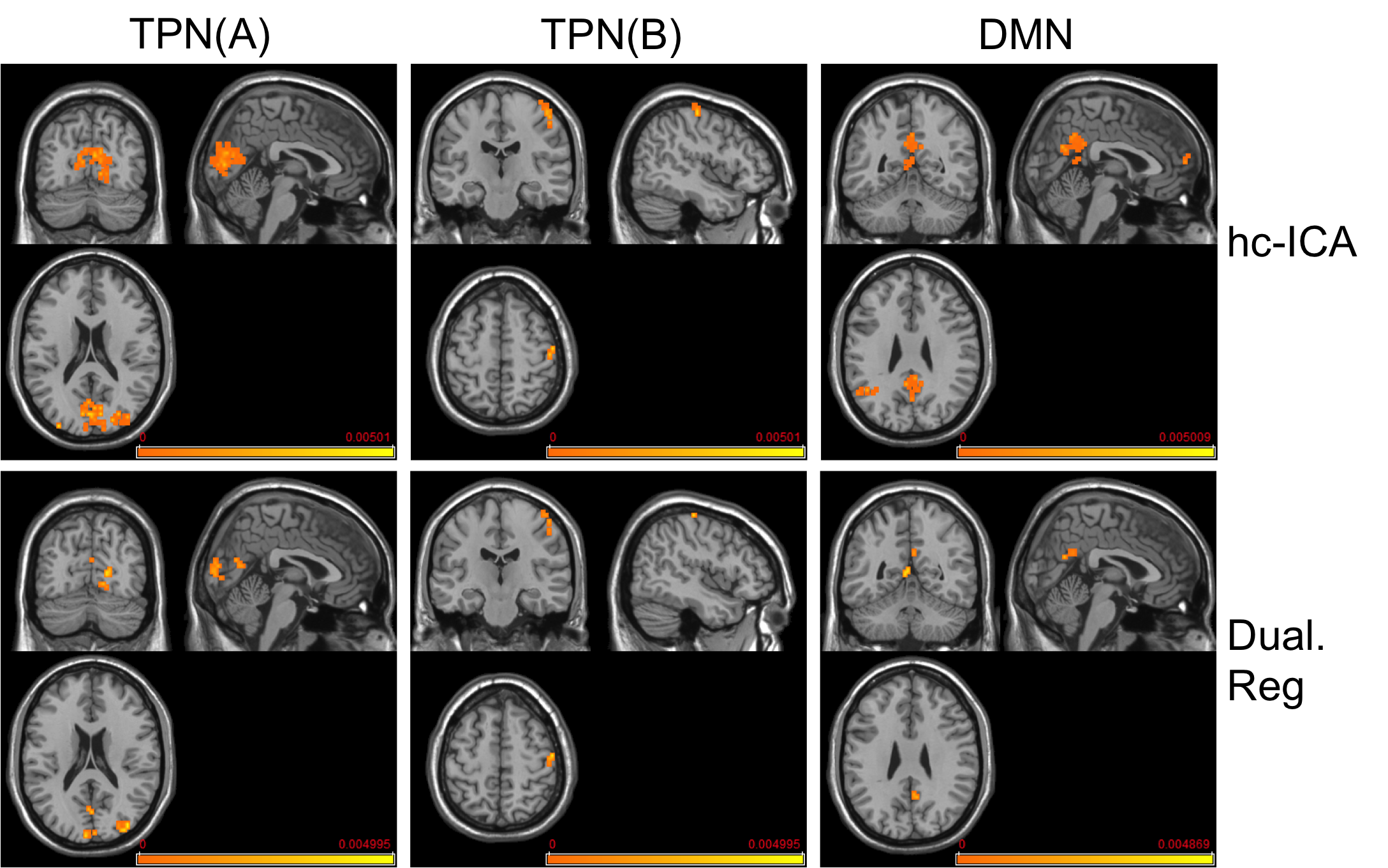}\\
		p\mbox{-values}\\
		\end{array}$
		\caption{The estimated covariate effects from hc-ICA and the dual-regression: the top panel shows the between group difference estimates within the two networks (meditator group minus control group); the bottom panel provides the $p$-values (Wald-type tests for hc-ICA, standard permutation tests for the dual-regression). All images thresholded at the corresponding $p$-values smaller than $0.005$}\label{fig3}
	\end{center}
\end{figure}

Figure \ref{fig3} presents the hc-ICA model-based estimates of between-group differences in the two networks and the associated $p$-values derived from the proposed inference procedure. In the TRN, meditation group demonstrated significantly stronger signals in both the visual processing region (VC) and the left hand region (HRSC). Since the tasks involved responding to visual stimuli via button clicking with the left hand, these two brain areas represented the key functional regions related to the experiment. The between-group comparison results indicated that the TRN of meditators demonstrated stronger functional connectivity, or better synergy, among its key regions as compared with control subjects.

We also examined the temporal correlations between the subject-specific time courses associated with the TRN and the experimental task time series (see Figure 1 in the supplementary materials for subject-specific time courses). The temporal correlations were $0.494\pm 0.113$ among the control and $0.601\pm0.123$ among the meditators. This result suggests that the temporal dynamics of the meditators' TRN were better registered to the experimental tasks, indicating that the meditators exhibited better capacity to regulate automatic conceptual processing in response to semantic stimuli.

In the DMN, the meditators displayed significantly stronger signals in the PCC and MPC. These two regions, particularly the PCC (the central node of DMN), have been found to play an essential role in the DMN \citep{greicius2003functional, leech2011fractionating}. Our results impled that these two key DMN regions had stronger functional connectivity, which indicated more coherent synergy, among meditators. We also found stronger signals in the LLPC for the meditators. This region is known to be associated with language processing and often becomes a more prominent subregion in the DMN, i.e. demonstrating stronger connectivity with the other regions within DMN, when the experimental stimuli are language-based. Our findings suggested that compared with control subjects, this language-related subregion in DMN is more functionally connected with the central regions of the DMN for the meditators. These facts all indicated that meditators had stronger functional connectivity within DMN than the control.

We compared our test results with the between-group differences estimated by the dual-regression ICA method whose $p$-values were calculated from randomized permutation tests \citep{beckmann2009group, filippini2009distinct} (Figure \ref{fig3}). In the TRN, the dual-regression method found much less between-group differences within the visual cortex and the left hand region. In the DMN, the dual-regression identified little distinction between the meditators and the control. Specifically, the dual-regression only showed a little between-group differences in the posterior cingulate cortex and it didn't detect any differences in the medial prefrontal cortex or the left lateral parietal cortex. These suggested that dual-regression failed to fully reveal the important distinctions between meditators and the control in the central node and the language processing node in DMN. Our proposed hc-ICA, however, is more powerful in detecting covariate effects on brain networks as compared with existing two-stage approaches such as the dual-regression method.

\begin{figure}[h]
\begin{center}
\small$
\begin{array}{ccc}
\mbox{TRN(A)} & \mbox{TRN(B)  } & \mbox{DMN}\\
 \includegraphics[scale=0.25]{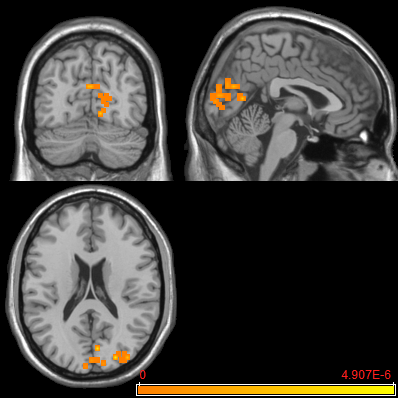}&
 \includegraphics[scale=0.25]{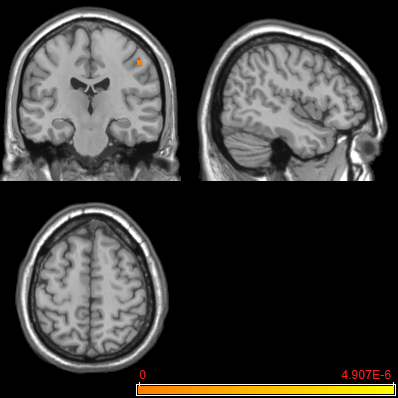}&
 \includegraphics[scale=0.25]{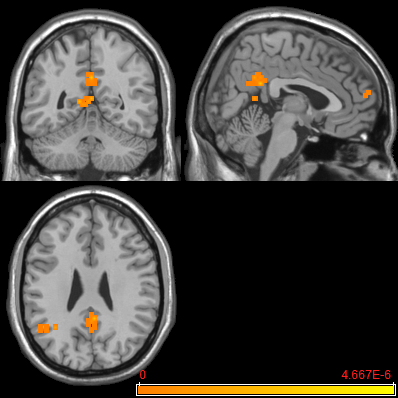}\\
\end{array}$
\caption{$p$-values at voxels with significant between-group differences after thresholding via the FDR control. Original $p$-value maps thresholded at FDR-corrected-$p<0.05$.}
\label{fig4}
\end{center}
\end{figure}

To adjust for multiple comparisons across voxels, we performed the procedure by \citet{benjamini2001control} to conduct FDR corrections on hc-ICA testing results within selected voxels (Figure \ref{fig4}). We obtained similar results and found that the meditators demonstrated significantly stronger signals in the key regions in the TRN and the DMN as compared with the control. The dual reg ICA didn't pick up any between-group differences in the two networks after the FDR correction.

\section{Discussion}\label{sec5}
We proposed a hierarchical covariate ICA (hc-ICA) to formally model and test covariate effects on brain function networks, which can potentially help advance understanding how subjects' demographic, clinical and biological characteristics affect brain networks. We develop a maximum likelihood estimation method based on EM algorithms for hc-ICA and also propose a statistical inference procedure to test covariate effects. Simulation studies show that our methods provide more accurate estimation and inferences for covariate effects on brain networks than the existing group ICA methods. Application of the hc-ICA to the Zen meditation fMRI study helps us obtain important findings regarding the differences in brain functional networks between experienced Zen meditators and controls.

One of the main challenges in statistical modeling of brain imaging data is the heavy computation load. In this paper, we develop computationally efficient estimation and inference procedures for the proposed hc-ICA model. In particular, by exploiting the sparsity in fMRI source signals, the subspace-based EM algorithm dramatically reduces the computational time for ICA via concentration on a subspace of the latent source states. We have shown both theoretically and empirically that the subspace-based approximate method is well-supported by the characteristics of fMRI signals and provides highly accurate results. The definition of the subspace implies that it corresponds to the case where there is little overlap in the spatial distributions of fMRI source signals, which is supported by the findings in the neuroscience literature. To further evaluate the performance of our method when there some deviation from this scenario, we have conducted additional simulation studies by generating spatially overlapping source signals. Our results show that even when there is small to moderate overlapping, the approximate EM still provides fairly accurate estimates for the ICs. The proposed subspace-based approximation method can potentially be generalized to other high-dimensional data sets with sparse signals when using finite mixture models.

Our method can be further extended to account for the spatially sparse structure of the covariate effects $\bbeta(v)$, i.e. covariates only affect a very small proportion of brain locations. One possible approach to account for the sparsity in covariate effects is to include regularization terms for $\bbeta(v)$ in the likelihood function to obtain shrinkage estimators.

\section*{Acknowledgements}
We thank Dr. Giuseppe Pagnoni for the Zen meditation data.

\begin{supplement}[id=suppA]
	\stitle{Supplementary materials to the paper ``Modeling Covariate Effects in Group Independent Component Analysis With Applications to Functional Magnetic Resonance Imaging"}	
	\slink[doi]{TBD}
	\sdescription{This document presents the following contents: the details about our EM algorithm such as the expressions of the Q-functions; the derivation of the conditional moments in the E-step and the updating rules in the M-step; the details about the approximation EM algorithm; the proof of Theorem  \ref{thm1}; the criteria of selecting activating voxels with in each  brain network; temporal mixing time series corresponding to the task related network for each subject; additional experiment results justifying our EM algorithm in real data analysis.}
\end{supplement}

\bibliographystyle{imsart-nameyear}
\bibliography{bib_hcICA}

\section*{Supplementary materials}
\subsection{The Q-functions in the E-step}
The Q-function in the E-step of our EM algorithms can be expressed as
\begin{equation}
Q(\Theta|\hTh^{(k)})=Q_1(\Theta\mid\hTh^{(k)})+Q_2(\Theta\mid\hTh^{(k)})+Q_2(\Theta\mid\hTh^{(k)})+Q_4(\Theta\mid\hTh^{(k)}),\nonumber
\end{equation}
where
\begin{align}
Q_1(\Theta\mid\hTh^{(k)}) &= -\frac{NV}{2}\log|\bm E|-\frac{1}{2}\sum_{v=1}^V\sum_{i=1}^N\mbox{tr}\bigg\{\bm E^{-1}\Big[\by_i(v)\by_i(v)'
-2\bA_iE[\bs_i(v)|\bby(v);\hTh^{(k)}]\by_i(v)' \nonumber\\
&+\bA_iE[\bs_i(v)\bs_i(v)'|\bby(v);\hTh^{(k)}]\bA_i' \Big]\bigg\},\nonumber
\end{align}
\begin{align}
Q_2(\Theta\mid\hTh^{(k)}) &= -\frac{NV}{2}\log|\bm D|-\frac{1}{2}\sum_{v=1}^V\sum_{i=1}^N\mbox{tr}\bigg\{\bm D^{-1}\Big[ E[\bs_i(v)\bs_i(v)'|\bby(v); \hTh^{(k)}]\qquad\quad\qquad\qquad\quad\nonumber\\
&+ E[\bs_0(v)\bs_0(v)'|\bby(v); \hTh^{(k)}] + \bbeta(v)'\bm x_i\bm x_i'\bbeta(v) - 2E[\bs_i(v)\bs_0(v)'|\bby(v); \hTh^{(k)}]\nonumber\\
&+ 2E[\bs_0(v) |\bby(v); \hTh^{(k)}]\bm x_i'\bbeta(v) - 2E[\bs_i(v)| \bby(v); \hTh^{(k)}]\bm x_i'\bbeta(v)    \Big]\bigg\},\nonumber
\end{align}
\begin{align}
Q_3(\Theta\mid\hTh^{(k)})
&= -\frac{1}{2}\sum_{v=1}^{V}\sum_{\ell=1}^q\sum_{j=1}^m p[{z_{\ell}(v)=j}|\bby(v);\hTh^{(k)}]\bigg\{\log\sigma^2_{\ell,j}+\frac{1}{\sigma^2_{\ell,j}}\Big[\mu_{\ell,j}^2\quad\quad\qquad\qquad \qquad\nonumber\\
&+ E[s_{0\ell}(v)^2|{z_{\ell}(v)=j}; \bby(v), \hTh^{(k)}]-2\mu_{\ell,j}E[s_{0\ell}(v)|{z_\ell(v)=j}, \bby(v); \hTh^{(k)}] \Big]\bigg\},\nonumber
\end{align}
\begin{equation}
Q_4(\Theta\mid\hTh^{(k)}) = \sum_{v=1}^{V}\sum_{\ell=1}^q\sum_{j=1}^mp[{z_\ell(v)=j}|\bby(v); \hTh^{(k)}] \log\pi_{\ell,j} ,\qquad\qquad\qquad\qquad\qquad\qquad\qquad\quad\nonumber
\end{equation}
and $\mathbf{y}(v) = [\by_1(v)',...,\by_N(v)']'$ contains all the observed data at voxel $v$ (for all the $N$ subjects). To evaluate the Q-functions, we need the joint conditional distribution, $p[\bbs(v), \bz(v) \mid \mathbf{y}(v); \Theta]$ where $\bbs(v)=[\bs_1(v)',...,\bs_N(v)', \bs_0(v)']'$.

\subsection{The derivation of conditional probabilities in the E-step}
In this section, we provide details the E-step in our exact EM. We mainly focus on deriving $p[\bbs(v), \bz(v) \mid \mathbf{y}(v); \Theta]$ as well as its marginals. By collapsing our model across the $N$ subjects as, for $v=1,...,V$,
\begin{equation}\label{fullmodel}
\mathbf{A}'\mathbf{y}(v)= \bm B \mathbf{x}+ \bU \bm\mu_{\bz(v)} + \bR \mathbf{r}_{\bz(v)} + \mathbf{e}(v),
\end{equation}
where $\mathbf{r}_{\bz(v)} = [\bm \gamma_1(v)',...,\bm \gamma_N(v)', \bm \psi_{\bz(v)}']'$ concatenates error terms in the second and third level models, $ \mathbf{e}(v)=[\be_1(v)',...,\be_N(v)']'$ contains random errors for the first level model across all subjects, $\mathbf{x} = [\bx_1',...,\bx_N']'$ represents all the covariate measurements, $\bm B = \bm I_N \otimes \bbeta(v)'$, $\bm U = \bm 1_N\otimes \bm I_q$, $\bm R = [\bm I_{Nq}, \bm 1_N\otimes \bm I_q]$ and $\mathbf{A} = \mbox{blockdiag}(\bm A_1,...,\bm A_N)$ is a combined mixing matrix with $\bm A_i$s as its block diagonal elements ($\mathbf{A}$ is also orthogonal). It is trivial to have that in \eqref{fullmodel}, $\mathbf{e}(v)\sim N(\bm 0, \bm\Upsilon)$ and $\mathbf{r}_{\bz(v)}\sim N(\bm 0, \bm \Gamma_{\bz(v)})$ where $\bm\Upsilon=\bm I_N\otimes \bm E$ and $\bm \Gamma_{\bz(v)} = \mbox{blockdiag}(\bm I_N\otimes\bm D, \bm \Sigma_{\bz(v)})$. Thus \eqref{fullmodel} can be represent as
$$\mathbf{y}_0(v)  \sim N(\bR \mathbf{r}_{\bz(v)}, \bm\Upsilon), \quad  \mathbf{r}_{\bz(v)}\sim N(\bm 0, \bm \Gamma_{\bz(v)})$$
where $\mathbf{y}_0(v) = \mathbf{A}'\mathbf{y}(v)-\bm B \mathbf{x}- \bU \bm\mu_{\bz(v)}$. This representation is a canonical Bayesian general linear model given $\bz(v)$. Then given $\bz(v)$ and conditional on $\bby(v)$, $p[\mathbf{r}_{\bz(v)}\mid \bby(v), \bz(v); \Theta] =g( \bm\mu_{\mathbf{r}(v)|\bby(v)}, \bm \Sigma_{\mathbf{r}(v)|\bby(v)})$ where
\begin{eqnarray}
\bm\mu_{\mathbf{r}(v)|\bby(v)} &=& \bm \Sigma_{\mathbf{r}(v)|\bby(v)}\bR'\bm\Upsilon^{-1}[\mathbf{A}'\mathbf{y}(v)-\bm B \mathbf{x} - \bU \bm\mu_{\bz(v)}],\nonumber\\
\bm \Sigma_{\mathbf{r}(v)|\bby(v)} &=& \left(\bR'\bm\Upsilon^{-1}\bR + \bm \Gamma_{\bz(v)}^{-1}\right)^{-1}.\nonumber
\end{eqnarray}
It is trivial to show that $\bbs(v) = \bm P\mathbf{r}_{\bz(v)} + \bm Q_{\bz(v)},$ where
$$\bm P=\left(\begin{matrix}
\bm I_{Nq}, & \bU\\
\bm 0, &\bm I_q
\end{matrix}\right), \quad
\bm Q_{\bz(v)}=\left(
\begin{matrix}
\bm B \mathbf{x}  + \bU\bm\mu_{\bz(v)}\\
\bm\mu_{\bz(v)}
\end{matrix}
\right),
$$
we can easily have that:
\begin{equation}\label{postp1}p[\bbs(v) \mid \bby(v), \bz(v);\Theta] =g(\bm P\bm\mu_{\mathbf{r}(v)|\bby(v)} + \bm Q_{\bz(v)}, \bm P\bm \Sigma_{\mathbf{r}(v)|\bby(v)}\bm P').\end{equation}

Next we need to find $p[\bz(v)\mid \bby(v); \Theta]$. From \eqref{fullmodel}, we have that $p[\mathbf{A}'\mathbf{y}(v)\mid \bz(v)] =g( \bm B \mathbf{x}+ \bU \bm\mu_{\bz(v)}, \bR \bm\Gamma_{\bz(v)}\bR' + \bm \Upsilon)$. Notice that $p[\bz(v)]=\prod_{\ell=1}^q \pi_{\ell, z_\ell(v)}$ for all $v$, by simply applying the Bayes' theorem,
\begin{equation}\label{postp2}
p[\bz(v) \mid \bby(v); \Theta] = \frac{\left[\prod_{\ell=1}^q \pi_{\ell, z_\ell(v)}\right]g(\mathbf{A}'\mathbf{y}(v); \bm B \mathbf{x}+ \bU \bm\mu_{\bz(v)}, \bR\bm\Gamma_{\bz(v)}\bR' + \bm \Upsilon)}{\sum_{\bz(v)\in \mathcal R} \left[\prod_{\ell=1}^q \pi_{\ell, z_\ell(v)}\right] g(\mathbf{A}'\mathbf{y}(v); \bm B \mathbf{x}+ \bU \bm\mu_{\bz(v)}, \bR\bm\Gamma_{\bz(v)}\bR' + \bm \Upsilon)},\end{equation}
where $\mathcal R$ is the range of $\bz(v)=[z_1(v),...,z_q(v)]', z_\ell(v)=1,...,m$, which contains $m^q$ distinct vectors in $\mathbb{R}^q$.

Given this probability distributions, the moments in the Q-functions can be easily derived and they all have analytical forms.

\subsection{Details of our M-step}
In the M-step, we update the parameters within our model as follows:
\begin{itemize}
	\item Update $\bbeta(v)$: for $v=1,...,V$,
	\begin{align}\label{m1}
	\hat{\bbeta}(v)^{(k+1)} = \left(\sum_{i=1}^N \bm x_i\bm x_i'\right)^{-1}\sum_{i=1}^N\bigg\{ \bm x_{i} \left(E[\bs_i(v)' | \mathbf{y}(v);\hTh^{(k)}] - E[\bs_0(v)' | \mathbf{y}(v); \hTh^{(k)}]\right)\bigg\}.
	\end{align}
	\item Update $\bA_i$: for $i=1,...,N$, we let
	\begin{equation}\label{m2}
	\breve{\bm A}_i^{(k+1)} = \left\{ \sum_{v=1}^V \by_i(v)E[\bs_i(v) | \bby(v); \hTh^{(k)}] \right\}\left\{\sum_{v=1}^V E[\bs_i(v)\bs_i(v)' | \bby(v); \hTh^{(k)}]\right\}^{-1},
	\end{equation}
	and then update $\widehat{\bm A}_i^{(k+1)}=\mathcal{H}(\breve{\bm A}_i^{(k+1)} )$ where $\mathcal{H}(\cdot)$ is the orthogonalization transformation.
	\item Update $\bm E = \bm I_q\nu_0^2$ with:
	\begin{align}\label{m3}
	\hat{\nu}_0^{2(k+1)} = &\frac{1}{TNV}\sum_{v=1}^{V}\sum_{i=1}^{N}\bigg\{ \by_i(v)'\by_i(v) - 2\by_i(v)'\widehat{\bm A}_i^{(k+1)} E[\bs_i(v) | \bby(v); \hTh^{(k)}] \\ \nonumber
	& + \mbox{tr}\left[\widehat{\bm A}_i^{(k+1)\prime}\widehat{\bm A}_i^{(k+1)}E[\bs_i(v)\bs_i(v)' | \bby(v); \hTh^{(k)}]\right]\bigg\}.
	\end{align}
	\item Update $\bm D = \mbox{diag}(\nu_1^2,...,\nu_q^2)$: for $\ell=1,...,q$,
	\begin{align}\label{m4}
	\hat{\nu}_\ell^{2(k+1)} = & \frac{1}{NV}\sum_{v}^V\sum_{i=1}^N\bigg\{ E[s_{i\ell}(v)^2 | \bby(v); \hTh^{(k)}] + E[s_{0\ell}(v)^2 |\bby(v); \hTh^{(k)}] \\ \nonumber
	& - 2E[s_{i\ell}(v)s_\ell(v)|\bby(v); \hTh^{(k)} ] + \hat{\bbeta}_\ell(v)^{(k+1)\prime}\bm x_i\bm x_i' \hat{\bbeta}_\ell(v)^{(k+1)}\qquad\qquad\qquad\\ \nonumber
	& + 2\left(E[s_{0\ell}(v)|\bby(v); \hTh^{(k)}]- E[s_{i\ell}(v)|\bby(v); \hTh^{(k)}]\right)\bm x_i' \hat{\bbeta}_\ell(v)^{(k+1)}   \bigg\},
	\end{align}
	where $\hat{\bbeta}_\ell(v)^{(k+1)}$ is the $\ell$th column of $\hat{\bbeta}(v)^{(k+1)}$.
	\item Update $\pi_{\ell,j}$:
	\begin{equation}\label{m5}
	\hat \pi_{\ell, j}^{(k+1)} = \frac{1}{V}\sum_{v=1}^V p[{\bz_\ell(v)=j}|\bby(v); \hTh^{(k)}].
	\end{equation}
	\item Update $\mu_{\ell,j}$:
	\begin{equation}\label{m6}
	\hat \mu_{\ell,j}^{(k+1)} = \frac{\sum_{v=1}^V p[{\bz_\ell(v)=j}|\bby(v); \hTh^{(k)}]E[s_{0\ell}(v)|{\bz_\ell(v)=j}, \bby(v); \hTh^{(k)}]}{V\hat \pi_{\ell,j}^{(k+1)}}.
	\end{equation}
	\item Update $\sigma^2_{\ell,j}:$
	\begin{equation}\label{m7}
	\hat \sigma_{\ell,j}^{2(k+1)} = \frac{\sum_{v=1}^V p[{\bz_\ell(v)=j}|\bby(v); \hTh^{(k)}] E[s_{0\ell}(v)^2|{\bz_\ell(v)=j}, \bby(v); \hTh^{(k)}]}{V\hat \pi_{\ell, j}^{(k+1)}}-[\hat\mu_{\ell, j}^{(k+1)}]^2.
	\end{equation}
\end{itemize}
Here, $E[s_{0\ell}(v)\mid z_\ell(v)=j, \bby(v);\Theta]$, $E[s_{0\ell}(v)^2\mid{z_\ell(v)=j}, \bby(v);\Theta]$ and $p[{z_\ell(v)=j}\mid\bby(v);\Theta]$ are the marginal conditional moments and probability related to the $\ell$th IC. They are derived by summing across all the possible states of the other $q-1$ ICs as follows,
\begin{equation}\label{marginal-exact}
E[s_{0\ell}(v)\mid z_\ell(v)=j, \bby(v);\Theta]=\frac{\sum_{\bz(v)\in \mathcal R^{(\ell, j)}}p[\bz(v)\mid \bby(v);\Theta]E[s_{0\ell}(v)\mid \bby(v), \bz(v);\Theta]}{p[{z_\ell(v)=j}\mid\bby(v);\Theta]},
\end{equation}
\begin{equation}
p[{z_\ell(v)=j}\mid\bby(v);\Theta]=\sum_{\bz(v)\in \mathcal R^{(\ell, j)}}p[\bz(v)\mid \bby(v);\Theta].
\end{equation}
where $\mathcal{R}^{(\ell, j)}$ is defined as $\{\bz^r\in \mathcal R: z^r_\ell=j\}$ for all $\ell=1,..,q, j=1,...,m$.

\subsection{Proof of Theorem 1}
We prove Theorem 1 by introducing a lemma.
\begin{lemma}
	If the elements of $\bz(v)=[z_1(v),...,z_q(v)]'$ are independent with $p[z_\ell(v)=j]=\pi_{\ell, j}$ for $j=1,...,m, \ell=1,...,q$, then
	$\label{sps}p[\bz(v) \in R_0\cup R_1] = \mathcal{F}(\bm\kappa)$ where
	\begin{equation}
	\mathcal{F}(\bm\kappa)=\frac{1+\sum_{l=1}^q\kappa_\ell}{\prod_{l=1}^q (1+\kappa_\ell) },
	\end{equation}
	with $\bm\kappa=[\kappa_1,...,\kappa_q]'$ and $\kappa_\ell = p[z_{\ell}(v)\neq 1]/p[z_{\ell}(v)=1]$ for all $\ell=1,...,q$.
\end{lemma}
The parameters $\bm\kappa=[\kappa_1,...,\kappa_q]'$ can be interpreted as the odds for a random voxel of being activated/deactivated versus exhibiting background fluctuation in IC $\ell$. Lemma 1 indicates that the probability of interest, $p[\bz(v) \in R_0\cup R_1]$, depends on $\{\pi_{\ell, j}\}$ only through the odds. The proof of Lemma 1 is provided as follows.
\begin{proof}
	Let $\tau_{\ell,j}=\pi_{\ell,j}/\pi_{\ell,1}, j=2,...,m$, then $\kappa_\ell=\frac{p[z_{\ell}(v)\neq 1]}{p[z_{\ell}(v)=1]}=\sum_{j=2}^m \tau_{\ell,j}$. By definition $R_1\cap R_0 = \emptyset$ and $p[\bz(v)\in R_0] = \prod_{\ell=1}^q p[z_\ell(v)=1]=\prod_{\ell=1}^q \pi_{\ell, 1}$. For a given $\bz(v)\in R_1$, suppose $z_t(v)=j>1$ for $t=1,...,q$ and $z_{\ell\neq t}(v)=1$, then $p[\bz(v)]=\tau_{t,j}\prod_{\ell=1}^q \pi_{\ell, 1}$. This implies that
	$$p[\bz(v)\in R_1] = \left(\sum_{t=1}^q\sum_{j=2}^m \tau_{t,j}\right)\prod_{\ell=1}^q \pi_{\ell, 1}= \left(\sum_{\ell=1}^q \kappa_\ell \right)\prod_{\ell=1}^q \pi_{\ell, 1}.$$
	Also we have that $\sum_{j=1}^m \pi_{\ell, j}= 1$ for all $\ell=1,...,q$, then $\pi_{\ell, 1} + \pi_{\ell, 1}\sum_{j=2}^m\tau_{\ell, j}=(1+\kappa_\ell)\pi_{\ell, 1}=1$, which gives
	$\pi_{\ell, 1}=1/(1+\kappa_\ell)$. Thus
	\begin{align}
	p[\bz(v)\in R_0\cup R_1]& = p[\bz(v)\in R_0]+p[\bz(v)\in R_1] \nonumber\\
	&=\left(1+\sum_{\ell=1}^q \kappa_\ell \right)\prod_{\ell=1}^q \pi_{\ell, 1}\nonumber\\
	&=\frac{1+\sum_{l=1}^q\kappa_\ell}{\prod_{l=1}^q (1+\kappa_\ell)  }
	\end{align}
\end{proof}

Based on Lemma 1, we prove Theorem 1 in the following.
\begin{proof}
	We notice that \begin{equation}\kappa_\ell=\frac{p[z_{\ell}(v)\neq 1]}{p[z_{\ell}(v)=1]}=\frac{1-\pi_{\ell, 1}}{\pi_{\ell, 1}}.
	\end{equation}
	For all $0<\epsilon<1$, let $\delta=\frac{\sqrt{\epsilon}}{\sqrt{\epsilon}+q}\in(0, 1)$. Then if $\pi_{\ell, 1}>1-\delta$, i.e., $\pi_{\ell, 1}>\frac{q}{\sqrt{\epsilon}+q}$, we have that $0<\kappa_\ell<\frac{\delta}{1-\delta}$ for all $\ell=1,...,q$. Based on the Taylor expansion for $p[\bz(v)\in R_0\cup R_1]=\mathcal{F}(\bm\kappa)$ at $\bm\kappa=\bm 0$, $\exists 0<\kappa_\ell^0<\kappa_\ell$ for all $\ell=1,...,q$, such that
	\begin{align}
	p[\bz(v)\in R_0\cup R_1]&=\mathcal{F}(\bm 0)+\sum_{\ell=1}^q \frac{\partial \mathcal{F}}{\partial \kappa_\ell}\bigg|_{\kappa_\ell=\kappa_\ell^0}\kappa_\ell\nonumber\\
	& = 1 - \sum_{\ell=1}^q \frac{\sum_{j\neq\ell}\kappa^0_j}{\prod_{j\neq\ell}(1+\kappa_j^0)}\frac{1}{(1+\kappa^0_\ell)^2}\kappa_\ell
	\nonumber\\
	& > 1- \sum_{\ell=1}^q\sum_{j\neq\ell}\kappa_\ell^2\nonumber\\
	& > 1-\left(\frac{q\delta}{1-\delta}\right)^2\nonumber\\
	& = 1-\epsilon
	\end{align}
\end{proof}

\subsection{Remarks on the approximate EM}
In the approximate EM, the conditional distribution $\bz(v) \mid \bby(v)$ is determined by the probability masses
\begin{equation}\label{postpapp}\tilde p[\bz(v) \mid \bby(v), \Theta] =\left\{ \begin{array}{cl}\frac{\left[\prod_{\ell=1}^q \pi_{\ell, z_\ell(v)}\right]g(\mathbf{A}'\mathbf{y}(v); \bm B \mathbf{x}+ \bU \bm\mu_{\bz(v)}, \bm\Gamma_{\bz(v)}\bR' + \bm \Upsilon)}{\sum_{\bz(v)\in \mathcal{\widetilde R}} \left[\prod_{\ell=1}^q \pi_{\ell, z_\ell(v)}\right] g(\mathbf{A}'\mathbf{y}(v); \bm B \mathbf{x}+ \bU \bm\mu_{\bz(v)}, \bm\Gamma_{\bz(v)}\bR' + \bm \Upsilon)}, &  \bz(v)\in \mathcal{\widetilde R} \\
0, & \bz(v)\in \mathcal{R}\backslash \mathcal{\widetilde R}
\end{array}\right.\end{equation}
where $\mathcal{\widetilde R} = \mathcal{R}_0\cup \mathcal{R}_1$.
Thus we use a sparse vector of probability masses, with concentration of measures on the subset $\mathcal{\widetilde R} = \mathcal{R}_0\cup \mathcal{R}_1$, to approximate the exact conditional distribution of $\bz(v)$ given $\bby(v)$.
The follow-up evaluations of the conditional moments in the E-step only involves $\bz(v)\in \mathcal{\widetilde R}$. And the corresponding definition of $\mathcal{R}^{(\ell, j)}$ is adapted to $\mathcal{\widetilde R}^{(\ell, j)}=\{\bz^r\in \mathcal{\widetilde R}: z^r_\ell=j\}$.

\subsection{Thresholding the spatial maps based on the ML estimates for functional brain networks}
We threshold the estimated spatial maps to identify the activated/deactivated regions of the brain within certain functional network. This goal can be achieved naturally through our model estimation based on conditional probabilities. To be specific, if we assume that $z_\ell(v)=j$ indicates the $\ell$th component be activated at voxel $v$, then we can calculate $p[{z_\ell(v)=j}\mid\bby(v);\hTh]=\sum_{\bz(v)\in \mathcal R^{(\ell, j)}}p[\bz(v)\mid \bby(v);\hTh]$, where $\mathcal{R}^{(\ell, j)}$ is defined as $\{\bz^r\in \mathcal R: z^r_\ell=j\}$ for all $\ell=1,..,q, j=1,...,m$. This probability characterizes the state of voxel $v$ within network $\ell$. We can then obtain the spatial map for a functional network by thresholding $p[{z_\ell(v)=j}\mid\bby(v);\hTh]$ with a pre-specified probability.

\begin{figure}[!h]
	\begin{center}
		$\begin{array}{c}
		\includegraphics[scale=0.35]{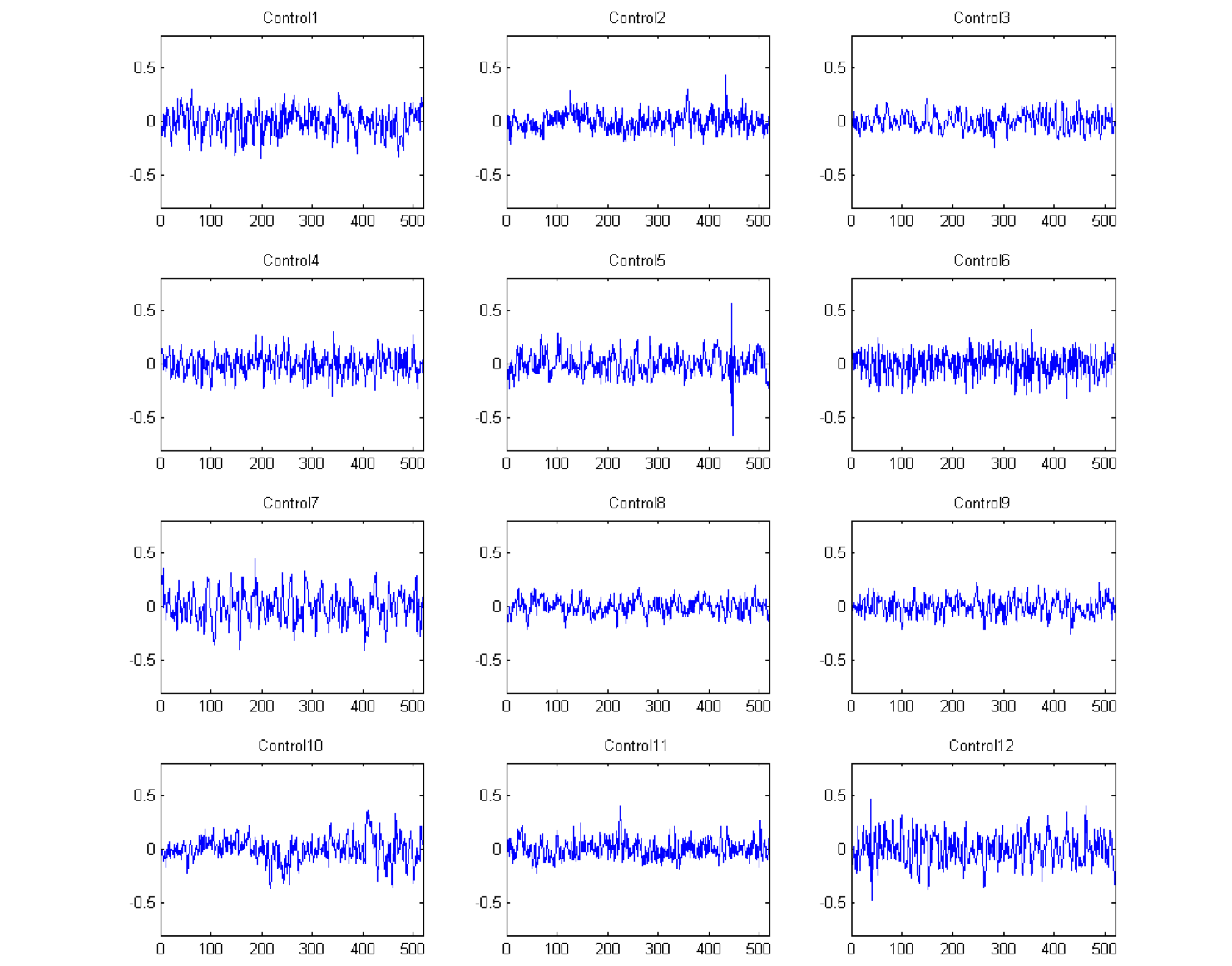}\\\hline
		\includegraphics[scale=0.35]{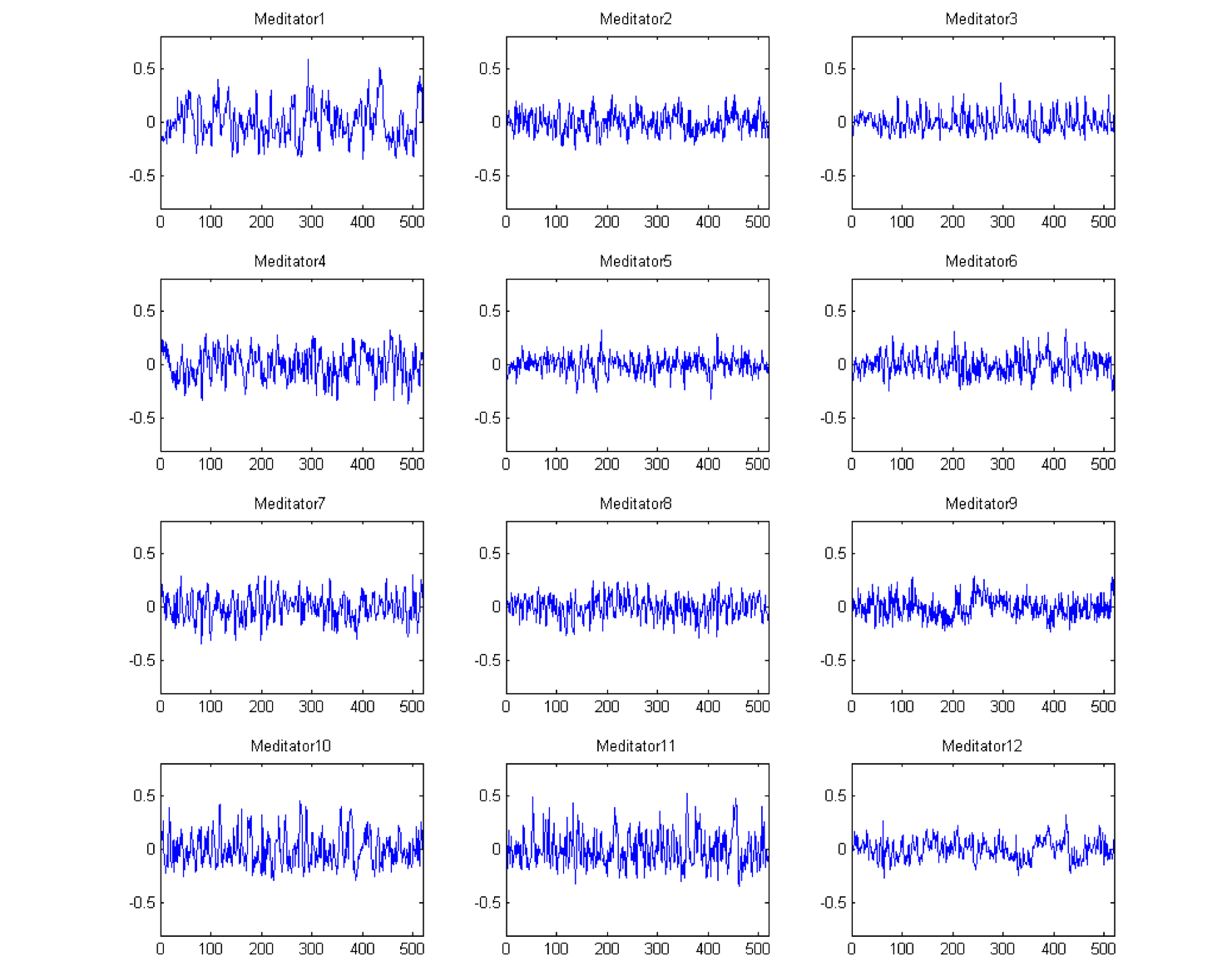}
		\end{array}$
		\caption{Subject-specific time series from the task-related network (correlation with the task time series: CTRL, $0.494\pm 0.113$; MEDT, $0.601\pm0.123$)}\label{fig1}
	\end{center}
\end{figure}
\subsection{The subject-specific time series for the task related network (TRN)}
We present the subject-specific time series for the TRN extracted from our hc-ICA method in Figure \ref{fig1}. The result shows that meditators present higher concordance with the tasks in terms of brain activities at the TRN compared to the control subjects.

\subsection{Checking the stability of our EM algorithm for the Meditation data analysis}
In addition the reported analysis, we repeated our implementation with another 6 sets of different initial values. We considered three scenarios with two experiments for each scenario: noise contaminated initial guesses (SNR =20), permutated initial guesses (randomly shuffling the order of initial guesses for the parameters of each network) and sign changed initial guesses (randomly changing the signs for the mixing matrices and the corresponding parameters). For the last two scenarios, we conducted necessary permutations and sign changes for the outputs of the algorithm. We then calculated the correlations of the estimated group-level spatial source signals, the mixing matrices and the covariate effects with our reported results. The correlations averaged across ICs are reported in Table \ref{emcov}. The results indicate that our algorithm is stable and provide strong evidence against local optima.
{\small
	\begin{table}[h!]
		\begin{center}
			\caption{Checking stability of the EM algorithm}\label{emcov}
			\begin{tabular}{cccccccccccc}
				\toprule\hline
				
				Experiment     &&Group ICs maps && Mixing matrices   && Covariate effects  \\ \hline
				Adding noises  && 0.961         &&  0.980            && 0.946    \\
				&& 0.978         &&  0.989            && 0.965    \\\hline
				Permutation    && 0.991         &&  0.993            && 0.984    \\
				&& 0.986         &&  0.993            && 0.980    \\\hline
				Changing signs && 0.935         &&  0.948            && 0.910    \\
				&& 0.928         &&  0.945            && 0.917    \\
				\hline\bottomrule
			\end{tabular}
		\end{center}
	\end{table}
}

\end{document}